# FMRI Clustering in AFNI: False Positive Rates Redux[†]

Robert W. Cox[*], Gang Chen, Daniel R. Glen, Richard C. Reynolds, Paul A. Taylor

*Scientific and Statistical Computing Core, NIMH/NIH/DHHS, Bethesda, MD, USA*

**Abstract**

Recent reports of inflated false positive rates (FPRs) in FMRI group analysis tools by Eklund et al. (2016) have become a large topic within (and outside) neuroimaging. They concluded that: existing parametric methods for determining statistically significant clusters had greatly inflated FPRs ("up to 70%," mainly due to the faulty assumption that the noise spatial autocorrelation function is Gaussian-shaped and stationary), calling into question potentially "countless" previous results; in contrast, nonparametric methods, such as their approach, accurately reflected nominal 5% FPRs. They also stated that AFNI showed "particularly high" FPRs compared to other software, largely due to a bug in 3dClustSim. We comment on these points using their own results and figures and by repeating some of their simulations. Briefly, while parametric methods show some FPR inflation in those tests (and assumptions of Gaussian-shaped spatial smoothness also appear to be generally incorrect), their emphasis on reporting the single worst result from thousands of simulation cases greatly exaggerated the scale of the problem. Importantly, FPR statistics depend on "task" paradigm and voxelwise $p$-value threshold; as such, we show how results of their study provide useful suggestions for FMRI study design and analysis, rather than simply a catastrophic downgrading of the field's earlier results. Regarding AFNI (which we maintain), 3dClustSim's bug-effect was greatly overstated‒their own results show that AFNI results were not "particularly" worse than others. We describe further updates in AFNI for characterizing spatial smoothness more appropriately (greatly reducing FPRs, though some remain >5%); additionally, we outline two newly implemented permutation/randomization-based approaches producing FPRs clustered much more tightly about 5% for voxelwise $p\leq0.01$.

---

## Introduction

Reports [1,2] of greatly inflated false positive rates (FPRs) for commonly used cluster-threshold based FMRI statistics packages (SPM, FSL, AFNI) caused a stir in both technical and semi-popular publications. In large part, this was due to dramatic summary statements about the collective previous approaches that, "Alarmingly, the parametric methods can give a very high degree of false positives (up to 70%, compared with the nominal 5%) for clusterwise inference," and[1], "These results question the validity of some 40,000 fMRI studies and may have a large impact on the interpretation of neuroimaging results" [2]. They hypothesized that a major factor for the problematic inflation was the mistaken assumption across the softwares that "spatial autocorrelation functions ... follow the assumed Gaussian shape." They contrasted these results with the output of nonparametric permutation tests, which they summarized simply as producing "nominal results for voxelwise as well as clusterwise inference." Additionally, Eklund et al. stated that a 15-year-old bug in 3dClustSim of the AFNI software package (which we maintain) resulted in "cluster extent thresholds that are much lower compared with SPM and FSL" and "particularly high FWE rates" [2] (FWE = family-wise error). Published responses to these findings ranged from saying that there is nothing new here ("tests based upon spatial extent become inexact at low thresholds" [4]), to cautious-but-concerned commentary [5], to inflated and conflated hyperbole about invalidating 15 years of research due to a software bug [6,7].

Here, we comment on several of these main points, not only by re-running simulations using some of the same "null" resting state FMRI data from the 1000 Functional Connectomes Project (FCON-1000 [8]), but also by examining closely the results from their own study (with paper [2] often referred in this text as "ENK16", for convenience). Eklund et al. did a vast amount of work in their study, running tens of thousands of simulations with several software packages, as well as making the data and scripts publicly available for further testing. Although they have raised an important issue that needs to be addressed, it must be noted that several of their extreme summarizing claims do not appear to be accurate characterizations of their own results.

The AFNI team takes the question of the inflated FPRs very seriously, but does not consider that the FMRI "apocalypse" has arrived. Rather than summarizing previous methods' results based on typical or average results, Eklund et al. repeatedly referred to the single worst result in over 3,000 simulations when summarizing parametric output ("up to 70% FPR"). We note that their own permutation method, under similar criterion, should be characterized as providing "up to 40% FPR" (both results come from ENK16-Fig. S9). While noting an upper limit of results may have some utility, in no case is that value a representative characterization of the method's typical performance. It would be much clearer and more informative to use median values and/or ranges to summarize and compare distributions of results; however, in ENK16, this was only done for their own nonparametric method.

---

[1] This second statement was later palliated to, "These results question the validity of a number of fMRI studies and may have a large impact on the interpretation of weakly significant neuroimaging results" [3].

This report comprises three themes. First, looking to the past, we repeat a subset of the Eklund et al. simulations using AFNI to show that the effect of the erstwhile bug within 3dClustSim, while not negligible, was not large and did not lead to "particularly high" FPRs. Looking at the present, we discuss modifications that have been made in clustering functionality with AFNI (as of 30 Jan 2017, version AFNI_17.0.03) in order to address the inflated FPRs in these tests. First, assumptions of spatial smoothness are evaluated and a new approach for estimating a non-Gaussian spatial autocorrelation function (ACF) is implemented, leading to greatly reduced FPRs in 3dttest++'s parametric approach, though in many cases still above the nominal 5%; then, we present a new permutation/randomization-based approach (generating cluster-size thresholds from the residuals at the group level) that is now implemented in AFNI, which shows FPRs clustered tightly about 5% across all voxelwise $p \leq 0.01$ thresholds. These methods, first outlined at the 2016 OHBM meeting [9], are demonstrated on sets of simulations following those in ENK16. Finally, looking to the future, an extension and generalization of this new approach to allow for spatially variable cluster-size thresholding shows promise in controlling FPRs while allowing for spatially variable smoothness in the FMRI noise. Throughout, we refer to various specific results of Eklund et al. from their main text, Supplementary Information, and tabular data they made available on GitHub. As in ENK16, all discussion of FPR refers to cluster counts at the whole brain level (as opposed to voxelwise *p*-values).

**Methods: Simulations**

For simulations performed here, we repeated the steps carried out in [1,2] using the 198 Beijing-Zang datasets from the FCON-1000 collection [8]. The detailed AFNI processing for individual subjects was somewhat different than in ENK16, since we ran with our most up-to-date recommendations for preprocessing (e.g., despiking, using 3dREMLfit with generalized least squares and ARMA(1,1) prewhitening to allow for temporal correlation, AnatICOR for denoising, and 3dQwarp for nonlinear registration; see the Supplementary Information for the complete alignment and afni_proc.py commands), but the results are quite comparable. The group analyses, using 1000 random sub-collections of the resting state FMRI Beijing datasets, were carried out using 3dttest++ in the same way as described in ENK16.

In ENK16, various combinations of simulation parameters produced widely varying levels of agreement or disagreement with the nominal 5% setting for FPRs for all software tools tested. To present a broad description, the comparisons presented here are for the set of basic scenarios put forth in ENK16. This includes investigating the four values of Gaussian smoothing applied (Full-Width at Half-Maximum (FWHM) of 4, 6, 8 and 10 mm) and two different voxelwise $p$-value thresholds (0.01 and 0.001; here, we also include the intermediate $p=0.005$). Four separate pseudo-stimulus timings were used in analysis of these null (resting-state) FMRI datasets: blocks of 10 s ON/OFF ("B1") and 30 s ON/OFF ("B2"), and event-related paradigms of (regular) 2 s task with 6 s rest ("E1") and (random) 1-4 s task with 6 s rest ("E2"). Each subject had the same stimulus timing for each case (as in ENK16).

**Results**

*The Past, I: 3dClustSim and "The Bug"*

One problem with past results was particular to AFNI: there was a bug in 3dClustSim. This program works by generating a 3D grid of independent and identically distributed $N(0,1)$ random deviates, smoothing them to the level estimated from the residuals of the FMRI data model at the individual level, carrying out voxelwise thresholding, and finally clustering to determine the rate at which contiguous clumps of different sizes occur at the various voxelwise thresholds. The bug, pointed out by the ENK16 authors in an email, was a flaw in how 3dClustSim rescaled the simulated 3D noise grid after smoothing in order to bring the variance of the values back to 1.0 (for ease of later $p$-value thresholding). This rescaling was off due to improper allowance for edge effects (effectively, zeros "outside" the grid were included in the smoothing), with the result being that the cluster-size thresholds computed were slightly too small, so that the FPR would end up somewhat inflated. During part of the work leading to [1,2], this bug was fixed in May 2015, and noted in the regular and publicly available log of AFNI software changes:

> "12 May 2015, RW Cox, 3dClustSim, level 2 (MINOR), type 5 (MODIFY)
> Eliminate edge effects of smoothing by padding and unpadding
> Simulate extra-size volumes then smooth, then cut back to the desired
> volume size. Can use new '-nopad' option to try the old-fashioned
> method. (H/T to Anders Eklund and Tom Nichols.)"[2]

Results comparing the pre- and post-fix versions of the standard 3dClustSim ("buggy" and "fixed", respectively) are shown in Fig. 1 (first two columns), which presents FPRs from re-running the two-sample *t*-tests (40 subjects total per each of 1000 3D tests) of ENK16. Comparing column heights for identical parameters, the size of inflation due to the bug ($\Delta$FPR = FPR$_{buggy}$ - FPR$_{fixed}$) was modest, particularly for *p*=0.001 where the inflation was $\Delta$FPR < 1-2%. At the less stringent voxelwise *p*=0.01, where FPRs had been noticeably more inflated for most software packages, the difference was greatest for the largest smoothing (understandably, given the problem within the program), with approximately $\Delta$FPR < 3-5%. In each case, the difference between the "buggy" and "fixed" values was small compared to the estimated FPR, meaning the bug had only a relatively minor impact (contrary to the statements in ENK16). Similar magnitudes of changes were found with one-sample *t*-test simulations. (Some differences between the "buggy" results herein and those in ENK16 are likely due to the improved time series regression procedures that we used; however, we did not systematically investigate the magnitude or details of this potential effect.)

At *p*=0.001, the "fixed" results for the event-related stimulus timings are not far from the nominal 5% FPR (Fig. 1, lower panel); however, the corresponding "fixed" results for the block-design stimulus timings are still somewhat high. The *p*=0.01 results are all still far too high in the "fixed" column (Fig. 1, upper panel). These results are revisited in the discussion of smoothness estimation, below.

*The Past, II: Parametric method results in AFNI and other softwares*

Looking at ENK16-Fig. 1, as well as at their Supplementary Information (SI) Figures S1-12, it is difficult to see how any one set of parametric results (SPM, FLS-OLS[3], AFNI-3dttest++ or AFNI-3dMEMA) can be classified as having a "particularly high" FPR compared to the others. This general similarity of FPRs is true even though the "buggy" 3dClustSim was used in creating them, further showing its small effect on inflation. For the purpose of comparison across softwares, Fig. 2 here shows boxplots of their combined results across all their simulation cases (publicly available on GitHub; see Appendix A here for further decomposition of the results). For voxelwise $p$=0.01, most parametric methods have results in the range of 15-40% FPR, and for $p$=0.001 most tests have 5-15% FPR. We also note that the oft-cited "up to 70% FPR" result was not produced by AFNI.

---

[2] https://afni.nimh.nih.gov/pub/dist/doc/program_help/history_all.html

The figure cited by Eklund et al. to demonstrate that AFNI's "cluster extent thresholds that are much lower compared with SPM and FSL" (ENK16-SI Appendix, Fig. S16) is not one based on general or summary results. Instead, it simply recapitulates the results of one set of simulations whose FPRs are already shown in one column of ENK16-Fig. S1-a, with two-sample comparison ($n$=10 subjects in each), one-sided testing, voxelwise $p$=0.01, and 6 mm smoothing for the "E2" (random event) task. Given the FPRs shown for that one case (AFNI ≈ 30% FPR, SPM ≈ 20% FPR, Perm ≈ 5% FPR), it is perhaps not surprising that AFNI's cluster extent in this case is smaller than SPM's by ≈ 50% and smaller than Perm's by a factor of five[4]. ENK16-Figs. S1-a and S16 show similar, consistent pieces of information for a single case, but this individual case is not a generic feature; certainly there are many simulations where AFNI had lower FPR and likewise a greater cluster extent. As noted above, a single case from a host of simulations does not provide grounds for representative characterization or generalization.

Additionally, we note that the comparison of cluster extents in ENK16-Fig. S16 did not account for neighborhood differences across the softwares. The examined software packages use different voxel neighborhood definitions by default: AFNI's 3dClustSim as run by Eklund et al. used a definition of nearest neighbor NN=1 (facewise bordering; n=6 voxels, but the user could set it to any value‐and now 3dClustSim always gives tables of results for NN=1, 2, and 3); SPM uses NN=2 (face+edge; $n$=18 voxels); FSL and Perm use NN=3 (face+edge+corner; $n$=26 voxels). The nearest neighbor definition should not in itself significantly affect final FPR values, as long as comparisons are made self-consistently between threshold and cluster volume within each package. However, any comparison of threshold cluster volumes across softwares would have to take this into consideration, as the volume differences are typically in a range of roughly 5-20% (it is unclear what additional information the cluster extents would add to the FPR comparisons themselves, once the neighborhood choices are accounted for).

---

[3] Here and below, we do not compare to FSL-FLAME1 results, which often had much lower FPRs for voxelwise p=0.01, and quite conservative results below the nominal FPR for $p$=0.001; further, we do not further investigate the other software tools, and merely note that FSL-FLAME1 had much different behavior from the rest of the "previous" approaches examined in [2].

[4] It is likely also consistent with FSL-OLS's result, but we do not know why FSL-FLAME1 is so much more conservative, assuming the same cluster size; Eklund et al. discuss this in [2].

Additionally, the parametric methods of AFNI, SPM and FSL were evaluated with an "ad hoc" clustering method from Cunningham and Lieberman (2009) [10], with results shown in ENK16-Fig. 2; "Perm" method results were not provided for this test. We note that we have *never* endorsed using this "ad hoc" approach, as it seems neither theoretically grounded nor experimentally generalizable. The initial Cunningham and Lieberman paper [10] applied a voxelwise $p$=0.005 with an additional extent threshold of +10 voxels to mimic the behavior of FPR=5%, as compared to running 1 million simulations; this was evaluated in only a single group of 32 subjects, where the additional value of "+10 voxels" was applied to a data set having 3.5x3.5x5 mm$^3$ voxels (for a total of 39,828 voxels within the whole brain mask) and 6 mm of blurring. Regardless of the results in that single data set, simply using a 10 voxel cluster threshold with any voxelwise $p$-threshold value in a data set of any arbitrary resolution and any processing stream cannot reasonably be expected to produce an appropriate FPR (e.g., ENK16 upsampled the FMRI data to 2x2x2 mm$^3$ voxels but still applied the same "+10 voxel" rule‐which in this case would have the volume of only 1.3 voxels from [10]). The fact that all presented software tools performed similarly in this test‐and with such poor apparent FPRs‐is unsurprising[5]. This particular clustering method is "ad hoc" *at best*, and absolutely unfounded when carried out under differing conditions (particularly spatial resampling).

In summary, we simply cannot understand the conclusion from Eklund et al.'s Results section that AFNI's FPR was "particularly high" compared to other software. The plotting of their own results (ENK16-Fig. 2) shows similar performance across all parametric software packages. This is more comprehensively observable in Fig. 2 here, which presents the medians and typical ranges of FPR values from all of the simulations performed in ENK16, and which provides a much better characterization of the results than using (some) extrema.

---

[5] Results of their own nonparametric approach with this ad hoc clustering method were not presented by ENK16, though it is difficult to see how they would be significantly different than the shown parametric methods.

*The Present, I: Updating "The Flaw" and assumptions about spatial smoothness*

The second problem in determining cluster-size threshold is much more widespread (to date) across the tools most used in the FMRI community: it is the flawed assumption that the shape of the ACF in the FMRI noise is Gaussian in form. That is, it has been generally assumed that, for voxels separated by Euclidean distance $r$, the spatial correlation between noise values has the form

(1) $\quad\quad f(r) = \exp[-r^2/(2b^2)], \quad\quad\quad$ with $b > 0$,

where it is traditional to specify the parameter $b$, and therefore the full shape of Eq. (1), by the related width parameter:

(2) $\quad\quad \text{FWHM} = 8\,[\ln(2)]^{½} \times b = 2.35482 \times b.$

In fact, as pointed out in ENK16, the empirical ACF shape (computed from the model residuals and averaged across the whole brain) has much longer tails than the Gaussian shape in Eq. (1). The heavy-tailed nature of spatial smoothness within the brain, which had been largely ignored previously, has significant consequences for thresholding clusters in FMRI analyses.

Fig. 3 illustrates the problem, along with the current solution adopted in AFNI. The empirical correlation falls off rapidly with $r$ at first, but then tails off much slower than the Gaussian function. We found that the empirical ACF estimates are typically well fit by a function that mixes the Gaussian and mono-exponential form

(3) $\quad\quad h(r) = a\,\exp[-r^2/(2b^2)] + (1-a)\,\exp[-r/c], \quad$ with $0 \leq a \leq 1$ and $b, c > 0$.

Given the demonstrated inadequacy of the pure Gaussian model in Eq. (1), 3dClustSim was modified to allow the generation of random 3D fields with autocorrelation given by Eq. (3). The mixed ACF model is now available in AFNI, and the (*a*,*b*,*c*) parameters are computed from each subject's time series regression model residuals in our FMRI processing stream tool (afni_proc.py).

To illustrate that the long tails in the spatial ACF can make a difference in cluster thresholding, we used 3dClustSim to compute the cluster-size threshold for a nominal 5% FPR using the 198 sets of estimated ACF parameters from the Beijing cohort. The voxelwise thresholds were taken as $p=0.010$ and 0.001 (left and right panels, respectively, of Fig. 4), from the NN=2 one-sided thresholding tables output by 3dClustSim. For each subject, 3dClustSim was run twice: once using a Gaussian ACF model with the FWHM estimated from the ACF mixed model (cf. Fig. 3), and once using the full mixed model ACF of Eq. (3). For $p=0.010$, the cluster-size thresholds estimated from the longer-tailed mixed model ACF are significantly larger than the Gaussian ACF model, showing that the long tails in the ACF have a large impact at larger $p$ thresholds. For $p=0.001$, the cluster-size thresholds from the two ACF cases differ far less, but are nontrivial. These facts are a major part of why the parametric software results in ENK16 are generally markedly "better" for the smaller $p$-value threshold (cf. Fig. 2).

FPRs from re-running the two-sample two-sided *t*-tests of ENK16, using this new "mixed ACF" model option in 3dClustSim, are also shown in Fig. 1 (third column). For the voxelwise threshold of $p=0.01$, the impact of the bug fix is much smaller than that of the long tail "fix" provided in the mixed ACF model. For the voxelwise threshold $p=0.001$, the impact of the bug fix is about the same as that of the long tails in the mixed ACF model, as these estimates were closer to the nominal 5% rate already. In each case the bug fix reduced the FPR values, as did the change to the mixed ACF model. Extensive further results, using several variations on this type of cluster analysis, with one- and two-sample, one- and two-sided *t*-tests, are given in Appendix B.

For the block designs (B1 and B2, 10 and 30 s blocks of "task"), the FPRs are still somewhat high even with the mixed ACF model. Preliminary investigations indicate this bias is partly due to the fact that the spatial smoothness of the FMRI noise is a function of temporal frequency‒that is, the FMRI noise at lower temporal frequencies is somewhat smoother than at higher frequencies. We do not know if that effect accounts for most of the disparity between the block and event-related designs.

*The Present, II: A nonparametric approach to cluster-size thresholding*

A second approach to adjusting the FPR in cluster-size thresholding has been implemented in the AFNI program 3dttest++ (which is also capable of incorporating between-subjects factors and covariates, in addition to carrying out the simple voxelwise *t*-tests implied by its name; perhaps it should have been named 3dOLStest). The procedure is straightforward:

- Compute the residuals of the model at each voxel at the group level;
- Generate a null distribution by randomizing among subjects the signs of the residuals in the test (and permuting subject datasets between groups, if doing a two-sample test without covariates), repeat the *t*-tests (with covariates, if present), and iterate 10,000 times;
- Take the 10,000 3D *t*-statistic maps from the randomization and use those as input to 3dClustSim (with no additional smoothing or random deviate generation): threshold the maps at a large number of *p*-values (e.g., 0.01, 0.005, 0.001, etc.), cluster-ize them, then record the maximum cluster size for each *p*-value from each *t*-map, accumulate statistics across the 10,000 *t*-maps, and thence determine the cluster-size threshold to achieve a given FPR, for each *p*-value.

All these steps are carried out by using the 3dttest++ program with the command line option '-Clustsim'.

The output is a table of cluster-size thresholds for a range of voxelwise *p*-value thresholds and a range of cluster-significance values. Such a table is produced for each of the clustering methods that AFNI supports: nearest neighbors NN=1,2,3, and one-sided or two-sided voxelwise thresholding. (In general, we prefer two-sided *t*-statistic thresholding in AFNI, as providing more transparency into the analysis results given the types of questions typically asked by researchers, but we do allow the user to opt for one-sided thresholding when appropriate[6].) These tables are saved in text format, and also stored in the header of the output statistics dataset for use in interactive thresholding in the AFNI GUI.

For comparison here, the 1000 two-sample *t*-tests described above were re-run for the 16 cases (4 blurring levels times 4 stimulus timings) with this new '-Clustsim' option, and tested against each of the 6 combinations of thresholding-sidedness and clustering-neighborliness possible in AFNI, over a range of voxelwise *p*-value thresholds. The results were similar across all 96 cases (and across sets of other tests, including one-sample, paired, and with covariates). The results for the one-sided *t*-test

---

[6] Performing a pair of one-sided tests in both directions under all circumstances would make it easier for a cluster to survive family-wise error correction, but it would also increase the FPR when a two-sided test is more appropriate. This unfortunate technique is all too common in FMRI.

NN=1 nearest neighbor clustering approach are shown graphically in Fig. 5; all FPRs are within the "nominal" 95% confidence interval for the FPR (3.65-6.35%) over the collection of voxelwise *p*-value thresholds tested. At this time, the use of this option is one of our recommendations, for cases where the group analysis can be carried out via a simple general linear model (GLM) with or without covariates.

*The Future, I: A third problem with cluster-threshold detection tools‒inhomogeneous smoothness*

A third standard assumption (present in AFNI, as well as the random field theory used in SPM and FSL [11]) also makes the idea of using a global cluster-size (or other cluster figure of merit) threshold somewhat non-optimal. The spatial smoothness of the FMRI noise is not spatially stationary‒it is significantly smoother in some brain regions (e.g., the precuneus and other large areas involved in the standard default mode network and also strongly affected by respiration artifacts) than in others; this inhomogeneity is also noted in [1,2]. The presence of variable smoothness means that the density of false positives for a fixed cluster-size threshold will differ across the brain, especially since the FPR is strongly nonlinear in the cluster-size threshold and in the noise smoothness. Using the same cluster-size threshold everywhere in such brain data will result in higher FPRs than expected in the smoother areas and lower FPRs than expected in the less-smooth areas.

A new AFNI program, 3dLocalACF, has been written to estimate the (*a*,*b*,*c*) parameters from Eq. (3) locally (in a ball, constrained within a brain mask) around each brain voxel. The non-Gaussian smoothness can be partly characterized by a new parameter called the "Full-Width at Quarter-Maximum" (FWQM), which characterizes the scale of the model in Eq. (3) at a broader point than the FWHM used in the simple Gaussian case; in the limiting case that the ACF is Gaussian, then FWQM = $2^{1/2}\times$FHWM. An example of the FWHM and FWQM smoothness estimates for one subject are shown in Fig. 6. We can only speculate as to the precise causes of this nonuniformity in spatial smoothness. It is possible that high-resolution FMRI methods can reduce this size of this problem [12].

A better approach to cluster-level detection must take into account this inhomogeneity. One approach would be to adaptively blur the data to make the spatial smoothness more homogeneous. An alternative approach is to adapt the cluster thresholding technique to deal with the spatial smoothness as it appears. It is this latter approach that we have chosen to develop first, through the randomization/permutation approach, as described in the next section. Another method, as yet unexplored, would be to generate synthetically non-stationary noise samples using the results from 3dLocalACF, and use those to compute spatially variable cluster-size threshold maps.

*The Future, II: Equitable Thresholding And Clustering (ETAC)*

Thresholding test statistics in realistic cases requires judgment, not just mathematics. Even in a simple case of testing a one-parameter alternative hypothesis (e.g., "mean = 0" vs. "mean ≠ 0"), one must judge between a one-sided test ("mean > 0") or two-sided ("mean < 0 or mean > 0"). In the latter case, one also has to judge whether it is necessary to give equal weight to both sides; that is, one could threshold so that under the null hypothesis of mean = 0, there is a 4% chance of a false positive (stating "mean > 0") and a 1% chance of a false negative (stating "mean < 0"). A principle of equity could reasonably force one to admit equal FPRs of either sign once the decision for a two-sided test has been made‑but this principle is not required by the mathematics or statistics of the situation (and could be contravened, given unbalanced external costs between the two types of outcomes).

Here, we define equity to mean that one treats equally situations that do not have important *a priori* relative differences. In this way, the number of arbitrary choices is reduced. For example, why choose between $p<0.01$ or $p<0.001$ (or some other semi-arbitrary $p$-value) as a voxelwise threshold? Such a fixed threshold approach may fail to detect a region that is anatomically small with high statistic values or anatomically large with low statistic values. It would be more desirable to set the voxelwise threshold within a range so that a "holistic" or equitable FPR can be achieved across various cluster sizes. Why require large cluster-size thresholds in brain regions that are not very smooth? It would be more equitable to use smaller cluster-size thresholds in less smooth brain regions and use larger cluster-size thresholds in more smooth regions.

A method for such balanced or equitable thresholding is under development in AFNI. It is an extension of the randomization of residuals approach described earlier, in the "-Clustsim" option of program 3dttest++. The new method produces a set of maps of the cluster-size threshold to use, one for each member of a range of voxelwise $p$-value thresholds. A voxel is "accepted" if it passes any one (or more) of the individual voxelwise $p$ plus cluster-size threshold tests. The cluster-threshold levels are chosen to be balanced, so that each cluster-threshold map (for one given $p$) contributes individually at the same FPR at each voxel, say α. In this way, no particular *p*-value and no particular location is specially favored: small high-intensity clusters are balanced with large low-intensity clusters; low-smoothness regions will get smaller cluster-size thresholds than high-smoothness regions. The global FPR is chosen by adjusting the individual map-wise α to get the desired final 5% rate. These calculations are implemented, using sign-randomization/permutation simulated *t*-test volumes generated via 3dttest++, in the new program 3dXClustSim, which generalizes the older 3dClustSim to derive voxel-specific cluster-size thresholds, as well as allowing the simultaneous use of several voxelwise *p*-thresholds.

Preliminary results from the Beijing subset (198 subjects) of the FCON-1000 collection are shown in Fig. 7. In this example, three *p*-value thresholds (0.005, 0.002, 0.001) are used simultaneously, and *p*-specific spatially variable cluster-size threshold maps are created from sign-randomized (and inter-sample permuted for the two-sample cases) simulations. Note that the calculated FPRs for each set of parameters (NN, sidedness, blurring, and stimulus) lie within the 95% confidence band of the nominal 5% value. Further ETAC analyses with 3dXClustSim are needed before it will be released for general use; some comparisons of intermediate, single *p*-value ETAC results with other methods are shown in Appendix B. Several generalizations are planned, and testing with task-based FMRI data collections will be carried out to compare the statistical power vs. alternative methods.

**Discussion and Conclusions**

*A note on "The Bug" and on bugs in general*

Of the many points considered here, we first commented on one of the least important and most publicly highlighted ones: the bug in the older versions of 3dClustSim (and its precursor AlphaSim). As shown here and noted before, this is actually a minor feature and the effect of the bug on FPR is relatively small. Correcting the underlying problem did indeed reduce the false positive rates in these tests, but the change in results cannot be considered a major factor in the overall FPR inflation. Both before and after the bug fix, 3dClustSim performed comparably to the other software tools being investigated. This is not to say that the presence of the bug was not unfortunate, but by itself, it could not "have a large impact on the interpretation" [2] of FMRI results. In order to make significant FPR changes in the results of 3dClustSim, new methods were required, which were also presented herein and which are discussed further below.

Reproducibility has been a major topic in the field of FMRI, with several proposals of "best practices" put forth in various forms. It is obvious that the presence of bugs in software (as well as misuses of software settings inappropriate in the context and incorrect methods implementations) damages the validity and reproducibility of reported results, and there is no greater concern for those writing software‑particularly when it is intended for public use‑than preventing bugs. Much of the discussion surrounding ENK16, particularly in comments to and take-aways chosen by the popular press, focused on the bug that was present in 3dClustSim. The discovery of this bug was highlighted and hyped as a major component for rejecting 15 years of brain studies and (up to) 40,000 peer-reviewed publications on the brain, under the tacit or explicit assumption that the reported results would be unreproducible.

However, rather than being evidence for "a crisis of reproducibility" within the field of FMRI, the advertisement of the bug is itself an important verification of the reproducibility of FMRI analysis. In this imperfect world, the philosophy for maintaining AFNI has always been to correct any bugs and to update the publicly available software as soon as practicable, often posting on the public Message Board for significant changes. The AFNI group maintains a permanent and public list of updates/changes/bugs online, which we view as an important resource for users and an aid for supporting reproducibility.

While certainly an annoying moment for the researchers who used 3dClustSim (and for those who maintain the software), the knowledge and dissemination of this bug is part of the reproducibility process. The existence of software bugs is unfortunate but inevitable. Even huge distributions such as Python, Windows, Mac, and Linux release bug fixes regularly. Clarity of description and speed of repair are the best tools for combating their effects once discovered.

*The state of clustering*

There were many valuable points raised in the work of ENK16. Several of these were important for general consideration within the FMRI field, such as the assumption of most clustering approaches that spatial smoothness was well-enough approximated by a Gaussian shape. To address this point, we have shown how an updated approach within AFNI using an estimated non-Gaussian ACF greatly improves the FPR controllability within the test datasets. Additionally, there is also a new nonparametric method for clustering within AFNI that shows promise; however, this type of approach in general currently appears to be limited by practical considerations (that hold across software implementations) to relatively basic group analyses that can be performed through univariate GLM.

A permutation and/or randomization approach seems able to provide proper FPR control with few apparent assumptions; so why not use this approach for everything in FMRI group analysis? Permutation tests of complex models (e.g., complex AN(C)OVA or LME) can become extremely computationally expensive, especially when coded in an interpreted language such as Matlab, Python, or R, where not all statistics are easily re-computable hundreds or thousands of times [13]. Nor is permutation/randomization easily implemented in complex situations with nesting and/or covariates. Issues of smoothness inhomogeneity of the noise structure in brain images present significant challenges for the development of a parametric method for spatial thresholding‑but such a method would be very useful. Considering the overall complexity of the problem, it appears unlikely that a "gold standard" *prima facie* correct method for spatial thresholding of neuroimaging data will appear soon.

Additionally, a permutation test is neither always required nor uniformly advantageous, as it may sacrifice power unnecessarily in some cases. For example, a fixed number of permutations would set a lower bound for the *p*-value that could be achieved (or require distributional extrapolation [14]), leading to failure to detect a small cluster with a potentially very high significance level that would survive through a parametric approach or a much larger number of permutations. While permutation testing may be useful and even necessary in some situations, a general rule for determining those cases is not clear, and, as noted above, it may be computationally or methodologically prohibitive to use (e.g., in the common case of including covariates, missing data, mixed effects, etc.). Further work is required on this important issue for the field.

*Final (for now) thoughts on statistics in FMRI‑and some recommendations*

Certainly, when using a clusterwise FPR value, one would hope that a method would reliably reflect the nominal rates. But, in conjunction with other trending discussions in the statistics literature, *p*-value thresholds are not sacred boundaries so that results around them live or die by tiny fractions above or below them; instead, thresholds are a convenience for focusing on reporting, but they are only part of the story. Our point here ties into discussions of reducing '*p*-hacking' and emphasizing effect sizes in results reporting for FMRI [15]. One beneficial "side effect" of the equitable clustering method, as shown in Fig. 7, is that results depend less sensitively on user-optional parameters such as blurring radius, NN value, and *p*-value(s) used for decision making.

Statistical testing and reporting is far from the end of a neuroscientific FMRI paper; in fact, these steps are just the technical prelude to the neuroscientific interpretation. At present, the conclusions of a study depend strongly on previous work and knowledge, rather than relying solely on statistical arguments from the current data alone. It is very hard to decide without close examination if a weakly "active" (or "connected") cluster in a brain map is actually critical to forming the paper's conclusions.

Despite being over 25 years since the first BOLD FMRI experiments were carried out, the mapping of human brain activity and connectivity is still evolving in both methodology and interpretation. Many nontrivial mathematical models have been designed to describe the diverse phenomena observed in neuroimaging research, yet most remain without "gold standard" verification. While improvements to methodology can help the situation (e.g., by increasing the reliability and specificity of characterization), authors, editors, and reviewers should continue to use statistical thresholding as a source of information, but not as the final authority.

One option for somewhat reducing the cluster size threshold in group analyses is to use accurate nonlinear alignment to a template, and then perform clustering within a slightly inflated gray matter mask instead of a whole brain mask. In the analyses shown herein, a whole brain mask comprising 74,397 3x3x3 mm$^3$ voxels was used. In a few simulations, we ran parallel analyses within a gray matter mask of 43,824 voxels in order to get a feel for the subsequent change in cluster size thresholds; on average, their sizes were reduced by about 25% (more at *p*=0.010, less at *p*=0.001). Since most FMRI-detectable activation or long-distance correlation occurs in gray matter, this approach has the potential for increasing statistical power without inflating FPR. It does depend strongly on the accuracy of the alignment to the template: affine or low order nonlinear alignment is *not* adequate for this purpose. Use of such a mask needs to be carefully vetted in any particular application, to ensure that all brain regions of potential interest are included, and that the subjects' alignments are all carried out to the needed precision.

At the time of writing, our recommendations for AFNI users are the following:

- When carrying out a group analysis via one- or two-sample testing, or GLM with between-subjects factors/covariates, one can have cluster-size thresholds determined by nonparametric analysis (using program 3dttest++ with the "-Clustsim" option); this has been tested extensively and gives very well-behaved FPR, as shown herein.
- Group analysis can also be carried out via Monte-Carlo simulations (using 3dClustSim) with the new mixed model ACF option ("-acf") to account for the noise smoothness structure; for instance, in the case of more complex models (e.g., Linear Mixed Effects [13], 3dLME), this likely must be the method used, at present, due to modeling limitations with GLM. The cluster parameters should be derived from the mean of the individual subject mixed model ACF parameters (which are computed by the standard afni_proc.py pipeline), and a voxelwise $p$=0.001 or $p$=0.002 is reasonably safe with this method, as in Appendix B (and compare, for example, our Fig. B-9 with ENK16-Fig. S2 SPM results).

In the near future, we hope to release the ETAC (3dXClustSim) method for general use with 3dttest++; however, further testing, development, and documentation are required before ETAC is fully ready.

**Author Disclosure Statement**

The authors declare no competing financial interests.

**Acknowledgements**

The authors thank Alfonso Nieto-Castanon for informative discussions and figure suggestions. The research and writing of the paper were supported by the NIMH and NINDS Intramural Research Programs (ZICMH002888) of the NIH/DHHS, USA. This work extensively utilized the computational resources of the NIH HPC Biowulf cluster (http://hpc.nih.gov).**References**

1. Eklund A, Nichols T, Knutsson H. Can parametric statistical methods be trusted for fMRI based

group studies? https://arxiv.org/abs/1511.01863 (2015).

2. Eklund A, Nichols T, Knutsson H. Cluster failure: Why fMRI inferences for spatial extent have inflated false-positive rates. PNAS 113: 7900-7905 http://www.pnas.org/content/113/28/7900.full (2016).

3. Nichols T. Errata for Cluster failure. http://blogs.warwick.ac.uk/nichols/entry/errata_for_cluster/ (2016).

4. Flandin G, Friston KJ. Analysis of family-wise error rates in statistical parametric mapping using random field theory. https://arxiv.org/abs/1606.08199 (2016).

5. Stockton N. Don't Be So Quick to Flush 15 Years of Brain Scan Studies. http://www.wired.com/2016/07/dont-quick-flush-15-years-brain-scan-studies/ (2016).

6. BEC Crew. A bug in FMRI software could invalidate 15 years of brain research. http://www.sciencealert.com/a-bug-in-fmri-software-could-invalidate-decades-of-brain-research-scientists-discover (2016).

7. Russon M-A. 15 years of brain research has been invalidated by a software bug, say Swedish scientists. http://www.ibtimes.co.uk/15-years-brain-research-has-been-invalidated-by-software-bug-say-swedish-scientists-1569651 (2016).

8. Biswal B, et al. Toward discovery science of human brain function. PNAS 107:4734–4739 (2010).

9. Cox RW, Reynolds RC. Improved Statistical Testing for FMRI Based Group Studies in AFNI :). OHBM Geneva (2016). https://afni.nimh.nih.gov/pub/dist/HBM2016/Cox_Poster_HBM2016.pdf

10. Lieberman MD, Cunningham WA. Type I and Type II error concerns in fMRI research: re-balancing the scale. Soc Cogn Affect Neurosci 4(4):423-8 (2009).

11. Worsley KJ, Marrett S, Neelin P, Vandal AC, Friston KJ, Evans AC. A unified statistical approach for determining significant signals in images of cerebral activation. Hum Brain Mapp, 4:58-73 (1996).

12. Wald LW and Polimeni JR. Impacting the effect of fMRI noise through hardware and acquisition choices – Implications for controlling false positive rates. NeuroImage (online ahead of publication) http://dx.doi.org/10.1016/j.neuroimage.2016.12.057

13. Chen GC, Saad ZS, Britton JC, Pine DS, Cox RW. Linear mixed-effects modeling approach to FMRI group analysis. NeuroImage 73:176-190 (2013).

14. Scholz FW. Nonparametric Tail Extrapolation. (White paper from Boeing Information & Support Services) http://www.stat.washington.edu/fritz/Reports/ISSTECH-95-014.pdf

15. Chen GC, Taylor PA, Cox RW. Is the Statistic Value All We Should Care about in Neuroimaging? NeuroImage, (in press) doi: 10.1016/j.neuroimage.2016.09.066 (2016).

**Appendix A. Additional parsing and plotting of Eklund et al.'s FPR results**

Fig. A-1 shows the original results of Eklund et al. (as in Fig. 2 in the main text here; made public in their GitHub repository) dissected into their results from one- and two-sample testing and by task stimulus (block designs B1 and B2, and event-related E1 and E2). Several useful comparisons of the effect of both the statistical test and stimulus paradigm are observable. For event-related stimuli, two-sample tests show uniformly lower FPR distributions (lower mean, lower max and fewer outliers), across software and for either voxelwise *p*. In particularly, for *p*=0.001, two-sample testing and event-related stimuli, all methods are clustered closely near the nominal 5% FPR. For the (shorter) B1 blocks, a reduction of FPR is also seen with two-sample testing, though with the (longer) B2 it is much less apparent. Again we note that, while there are differences in software results for the various parameters, the parametric methods tend to perform quite similarly for most subsets of parameters. The notable patterns of FPR results for a given *p*-value, statistical test and stimulus may be useful in guiding further studies.

**Appendix B. Studies with 3dttest++ and 6 different cluster-size thresholding methods**

Extensive null simulations were carried out on the 198 Beijing-Zang datasets from FCON-1000 collection. All the data (reorganized slightly from the original FCON-1000 download) and scripts used for the processing and analysis shown in this Appendix are available at the Globus shared Endpoint "ClusteringDataBeijing", stored on the NIH HPC Data Transfer server. This collection can be downloaded by establishing a free Globus user account, installing the free Globus client software, then searching for this Endpoint in the "Transfer Files" dialog. See https://www.globus.org for more information and to get started. All 198 subjects from the collection were processed successfully with the supplied scripts. The processing scripts are also available at https://afni.nimh.nih.gov/pub/dist/tgz/Scripts.Clustering.2017A.tgz (no data).

In brief, FPRs were estimated from 1000 3dttest++ runs, for:
- 4 stimuli (B1, B2, E1, E2);
- 4 blurs (4, 6, 8, 10 mm);
- 4 voxelwise $p$ thresholds (0.001, 0.002, 0.005, 0.010);
- One- and two-sided $t$-tests;
- One- and two-sample $t$-tests (20 subjects in each sample, pseudo-randomly selected);
- 6 updated cluster-size thresholding methods (each simulation used the same pseudo-random samples):
  - ETAC[7] (spatially variable cluster-size thresholds);
  - CS-RT-res (3dClustSim using the randomized/permuted $t$-tests from the $t$-test residuals);
  - CS-ACF-res (3dClustSim using the mixed model ACF parameters estimated from the $t$-test residuals);
  - CS-FWHM-res (3dClustSim using the Gaussian model ACF with the FWHM derived from the mixed model parameters estimated from the $t$-test residuals);
  - CS-ACF-sub (3dClustSim using the mean across the tested subjects of the mixed model ACF parameters estimated from the subject's time series residuals‒this is the method used in Fig. 1, column 3);

---

[7] The full ETAC method (as described in the main text and shown in Fig. 7) integrates results over a range of voxelwise $p$-values. Each figure presented in this Appendix shows ETAC results for a single $p$-value only, so that this intermediate step can be evaluated as well as compared with other existing approaches.

○ CS-FWHM-sub (3dClustSim using the mean across the tested subjects of the FWHM
　　　　　　estimated from the mixed model applied to the subject's time series residuals).
In total, 1536 (=4×4×4×2×2×6) FPR estimates were generated by these simulations. These results are shown in Figs. B-1 through B-16.

In general, the permutation/randomization approaches (ETAC, CT-RT-res) provide good FPR control at the desired 5% rate. Some cases are too conservative (e.g., $p$=0.010, one-sample, one-sided for the E2 stimulus), and some are too liberal (ETAC for $p$=0.010, one-sample, two-sided for various stimuli/blur combinations); these results reiterate the benefits of using lower voxelwise $p$-values, as well as the preference for two-sample testing (particularly in this simulation setup, where the resting state "null" is *known* to not be structure-free noise). But none of these cases for these two methods are dramatically wrong, and most are consistently accurate for a range of experimental designs, blurring, $p$-value, etc.

For the methods where a parametric (Gaussian or mixed model ACF) approach is used to calculate the cluster-size threshold, the results vary much more dramatically. It is clear that the Gaussian ACF is always more liberal than the mixed model ACF, as it should be since the two approaches have the same FWHM by construction and the mixed model allows for longer range correlations. Somewhat surprisingly, it appears to be generally true that using the subject mean mixed model ACF parameters is more conservative than using the ACF parameters estimated directly from the $t$-test residuals. For this method (CS-ACF-sub), the FPR estimates are "reasonable" for the smaller $p$-values (0.001 and 0.002), and for the larger smoothing levels (8 and 10 mm). These results form the basis for our recommendations at the end of the main text of this paper.

**FIGURES**

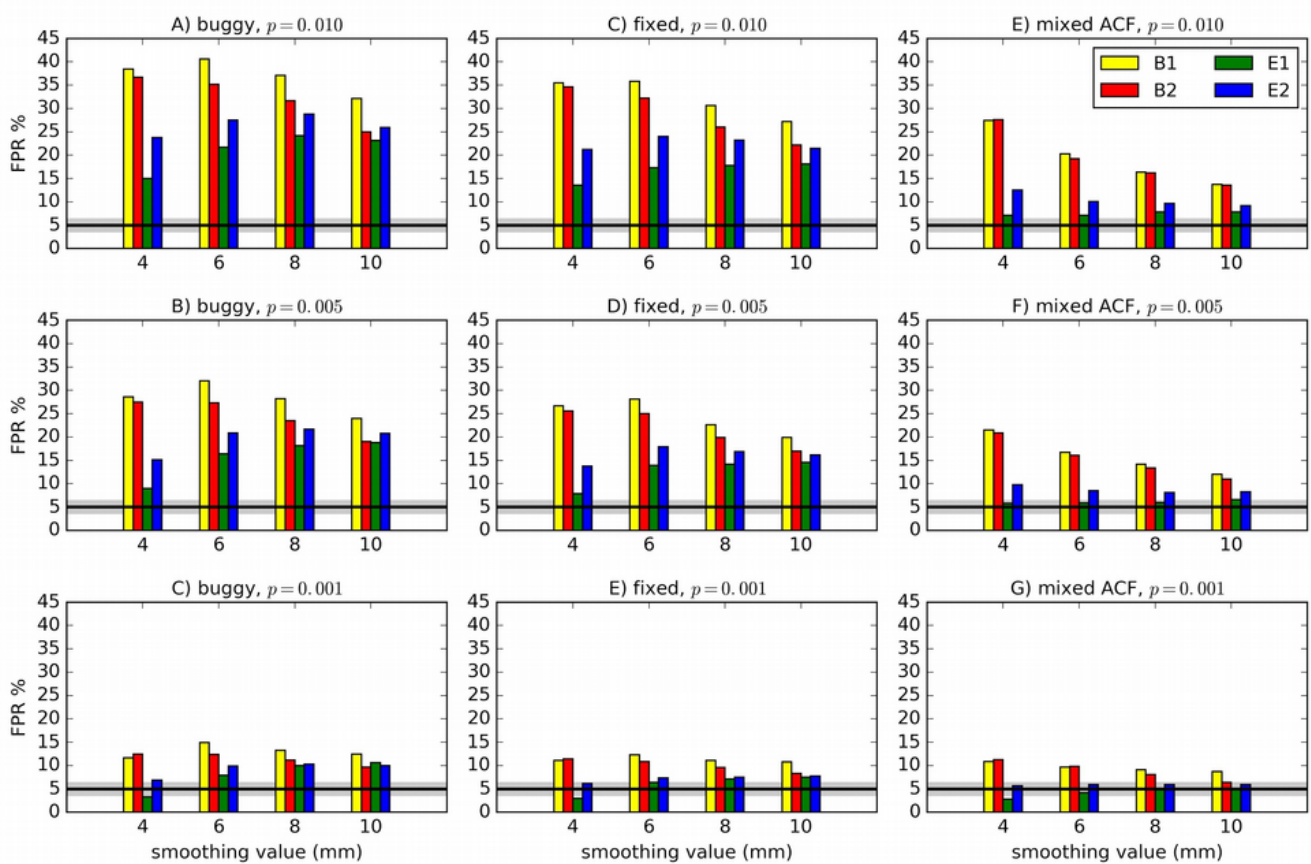

**Figure 1.** False Positive Rates (FPRs) for various software scenarios in AFNI, with 1000 two-sample 3D *t*-tests (as in [1,2]) using 20 subjects' data in each sample. "buggy" and "fixed" means that the cluster-size thresholds were selected using the Gaussian shape model with the FWHM being the median of the 40 individual subject's values: "buggy" and "fixed" via 3dClustSim before and after the bug fix, respectively. "mixed ACF" means that the cluster-size threshold was selected using Eq. (3) for spatial correlation of the noise, with the *a,b,c* parameters being the median of the 40 individual subject's values (estimated via program 3dFWHMx). Three different voxelwise *p*-value thresholds (one-sided tests, as used in [2]) are shown. The black line shows the nominal 5% FPR (out of 1000 trials), and the gray band shows its theoretical 95% confidence interval, 3.6-6.4%. As in ENK16, different smoothing values were tested (4-10 mm). B1 = 10 s block; B2 = 30 s block; E1 = regular event related; E2 = randomized event related.

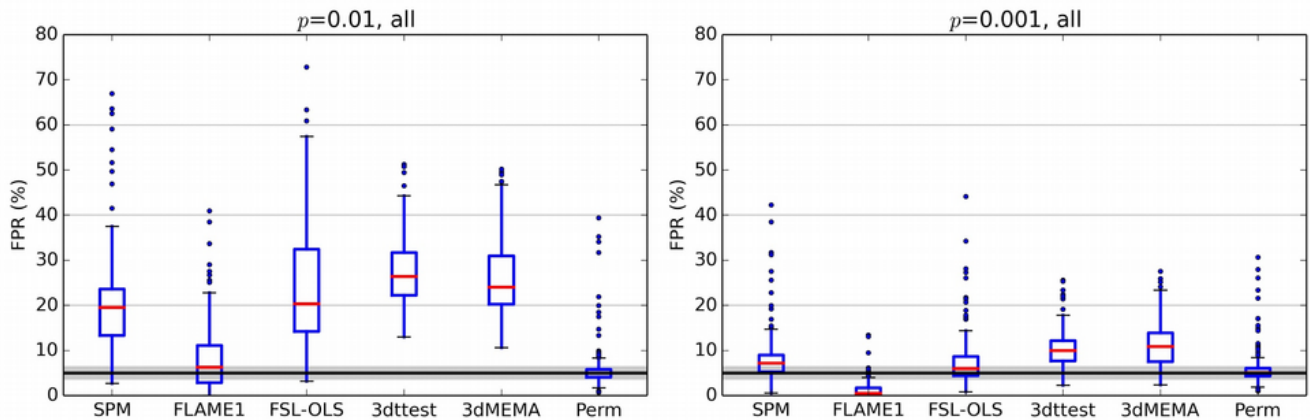

**Figure 2.** Summary of the FPR results examined in [2], combining all their test results (available from their GitHub repository). The results of each software across all voxelwise *p*=0.01 and *p*=0.001 cases are shown separately. Red lines show the median; the box covers the 25%-75% interquartile range; whiskers extend to the most extreme data point within 1.5x the interquartile range; and outliers are shown as dots. For a given voxelwise *p*, results are similar across parametric methods, with typical ranges of 15-30% FPR for *p*=0.01 and 5-15% FPR for *p*=0.001.

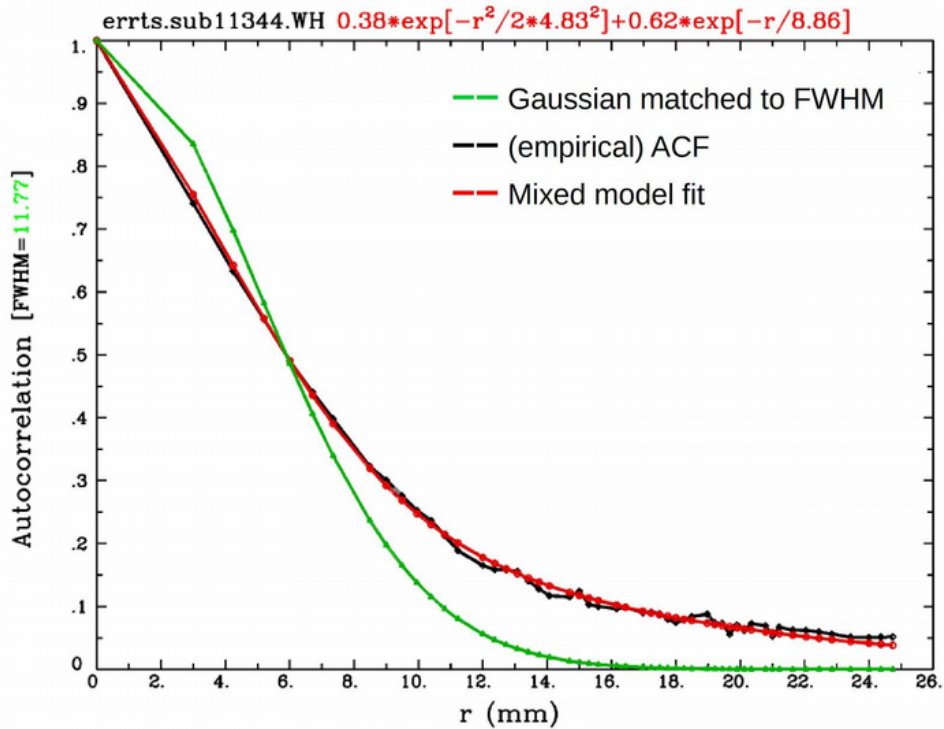

**Figure 3.** An example comparison of the original Gaussian fit (green) and the globally estimated empirical ACF values (black) from a single subject, which have large differences (importantly, in the tail drop-off above *r* ~ 8 mm). The proposed mixed model (red) after fitting parameters as described in Eq. (3) provides a much better fit of the data in this case (and in all cases in the datasets used herein). This plot is automatically generated in program 3dFWHMx.

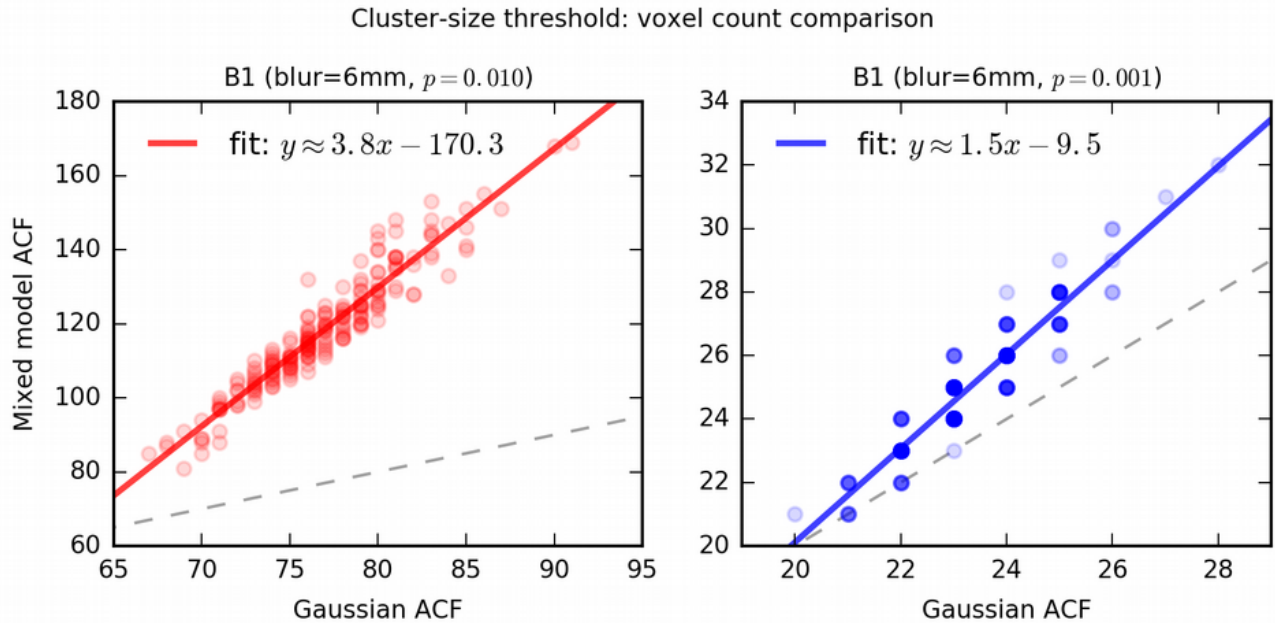

**Figure 4.** Cluster-size thresholds from 3dClustSim, ran over the estimated ACF for each of the 198 datasets in the Beijing-Zang cohort. The *x*-axis is the cluster-size threshold assuming a Gaussian-shaped ACF, with FWHM taken from the mixed model ACF estimate for each subject. The *y*-axis is the cluster-size threshold assuming the mixed model ACF of Eq. (3); the parameter estimates are computed from the residuals from the pseudo-stimulus B1, blur=6mm time series analyses. The left graph is for per-voxel *p* threshold 0.010; the right graph is for *p*=0.001. Approximate linear fits are shown overlaid; the dashed gray line shows *x=y*, providing a reference to indicate the disparity in cluster-size thresholds between the Gaussian and mixed-model ACF assumptions. Darker circles indicate points where multiple subjects had the same pair of thresholds (which are integer-valued). Cluster-size thresholds are taken from the NN=2, one-sided tests table output from 3dClustSim (which also output tables for NN=1 and NN=3, and for two-sided tests).

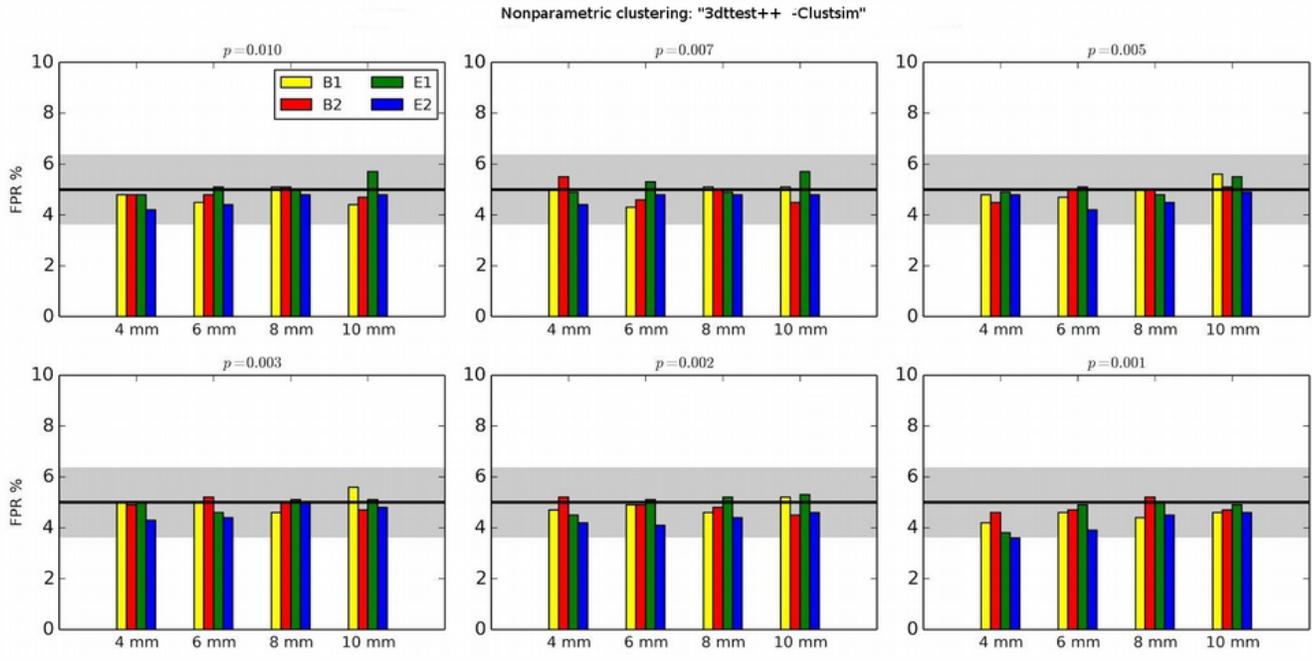

**Figure 5.** FPRs with cluster-size thresholds now determined from the '-Clustsim' option of 3dttest++ (one-sided tests with NN=1 clustering). See Fig. 1 for description of labels, but note that the *y*-axis range has been significantly changed here for visual clarity.

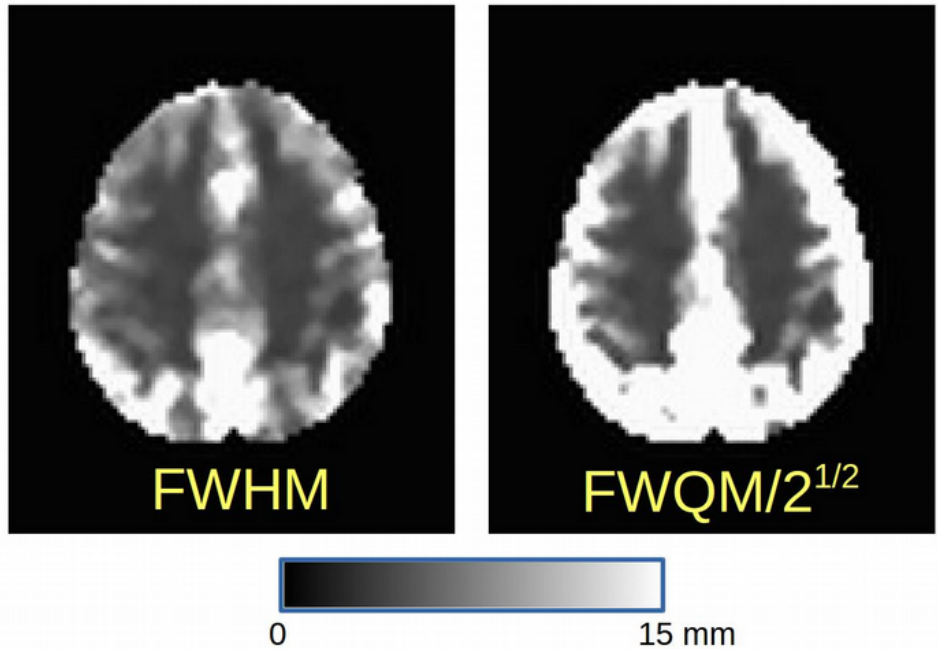

**Figure 6.** Images of the FMRI noise FWHM and the Full Width at Quarter Maximum (FWQM) from one subject (#11344) in the Beijing dataset collection (after nominal smoothing with a Gaussian kernel of 4 mm FWHM during preprocessing). The scale in both images is linear from black = 0 to white = 15 mm (and above). If the ACF were Gaussian, FWQM = $2^{1/2}$×FHWM. The FWHM map shows that the noise smoothness is not uniform in space (even within gray matter), and the FWQM map shows that the non-Gaussianity of the noise smoothness is also non-uniform. The magnitude of this effect on the FPR and how to allow for it in thresholding are still under investigation.

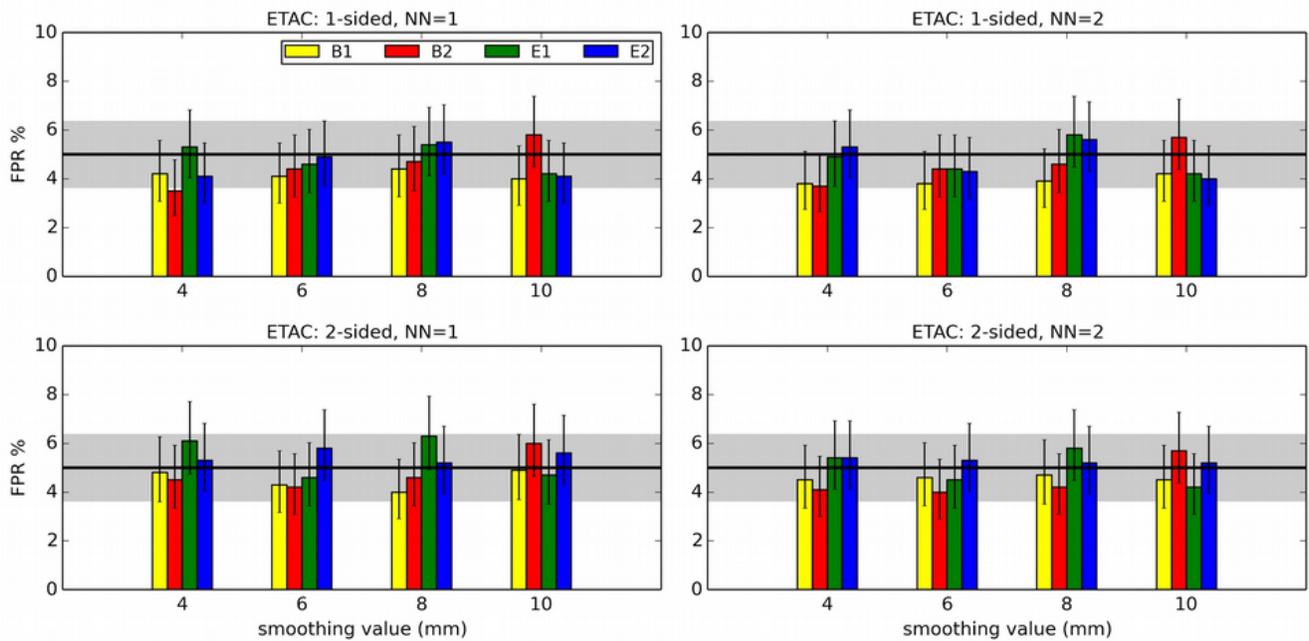

**Figure 7.** FPRs from the Equitable Thresholding And Clustering (ETAC) method, with the Beijing subset of FCON-1000. See Fig. 1 for description of labels, but note that the *y*-axis range has been changed here for visual clarity. Three *p*-value thresholds (0.005, 0.002, 0.001) are used simultaneously, and *p*-specific spatially variable cluster-size threshold maps are created from sign-randomized (and inter-sample permuted for the two-sample cases) simulations. For each of the 16 cases, 1000 random subsets of 40 subjects were selected, and a two-sample *t*-test was run between the first 20 and second 20 datasets for each of the 1000 instances. As labeled in each panel caption, results were calculated using either NN=1 or NN=2 neighborhoods (see text) to define the clusters, and either one-sided or two-sided t-testing to define the *p*-value thresholding. All FPRs fall within the 95% nominal confidence interval; error bars show the 95% confidence interval estimated for each result. FPRs from the Cambridge subset of the FCON-1000 (also 198 subjects) yielded similar results.

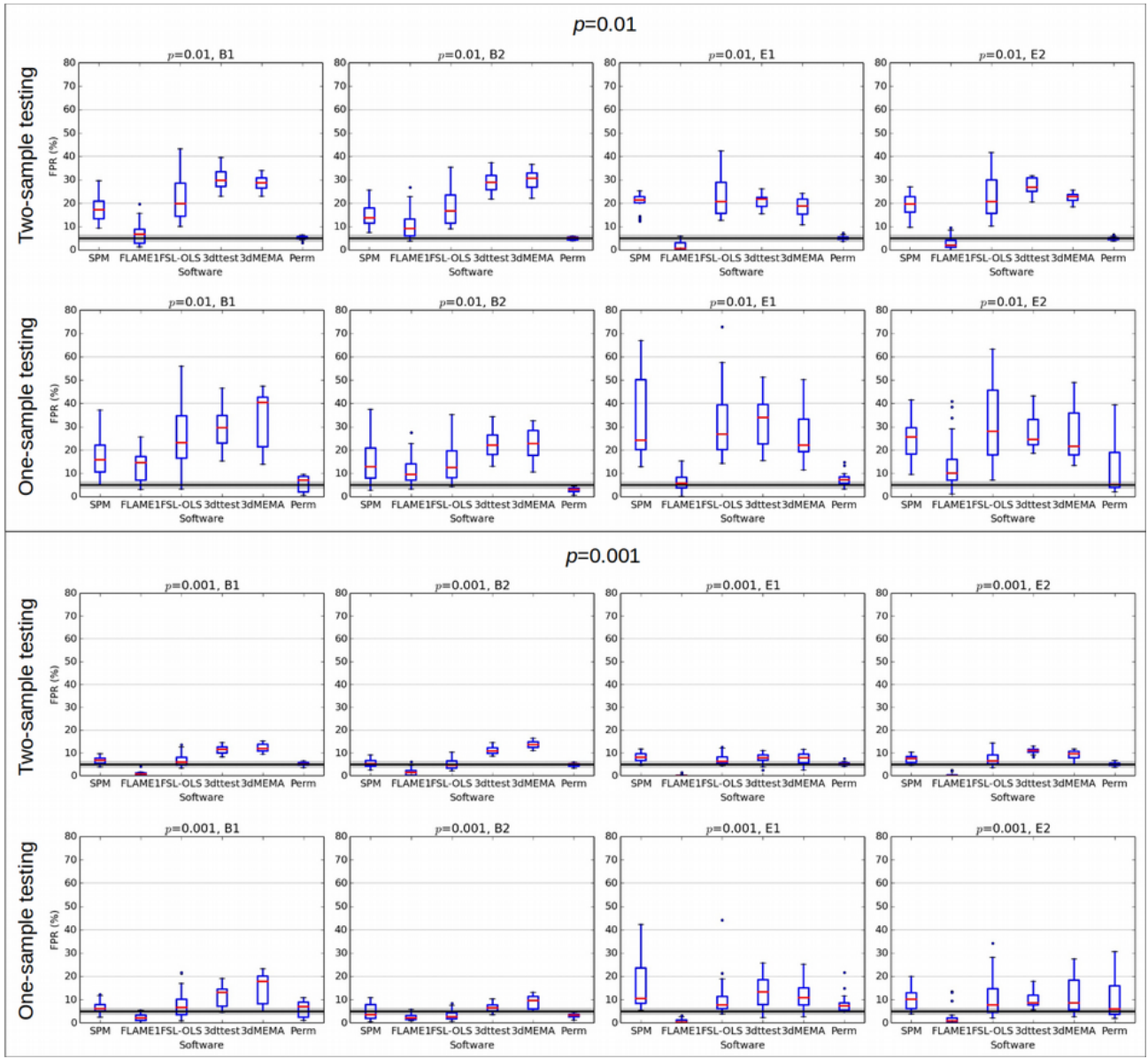

**Figure A-1.** Summary of the FPR results examined in ENK16, combining all their test results (available from their GitHub repository). The results of each software are shown separately based on voxelwise $p$ (=0.01 or 0.001), statistical test (one- or two-sample) and task stimulus (blocks B1 or B2, or event-related E1 or E2). Red lines show the median; the box covers the 25%-75% interquartile range; whiskers extend to the most extreme data point within 1.5x the interquartile range; and outliers are shown as dots. While results are fairly similar across parametric approaches, there is notable variation in FPR distribution among cases.

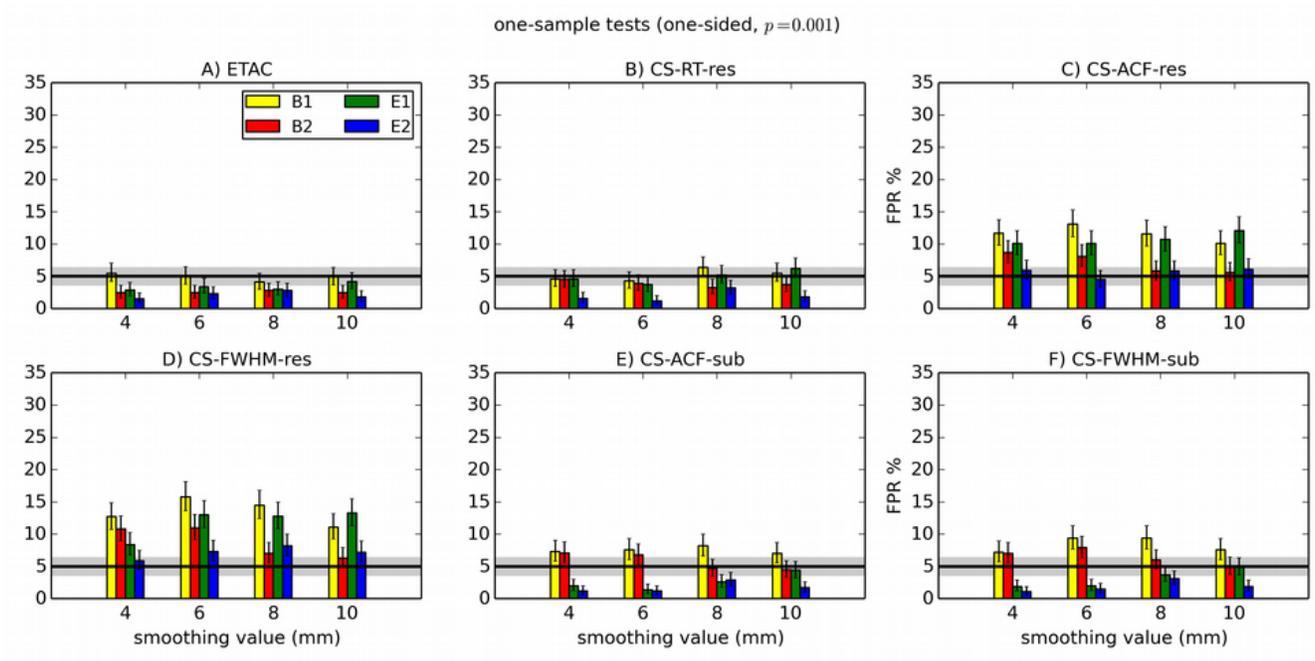

**Figure B-1**. One-sample, one-sided *t*-tests with voxelwise *p*=0.001.

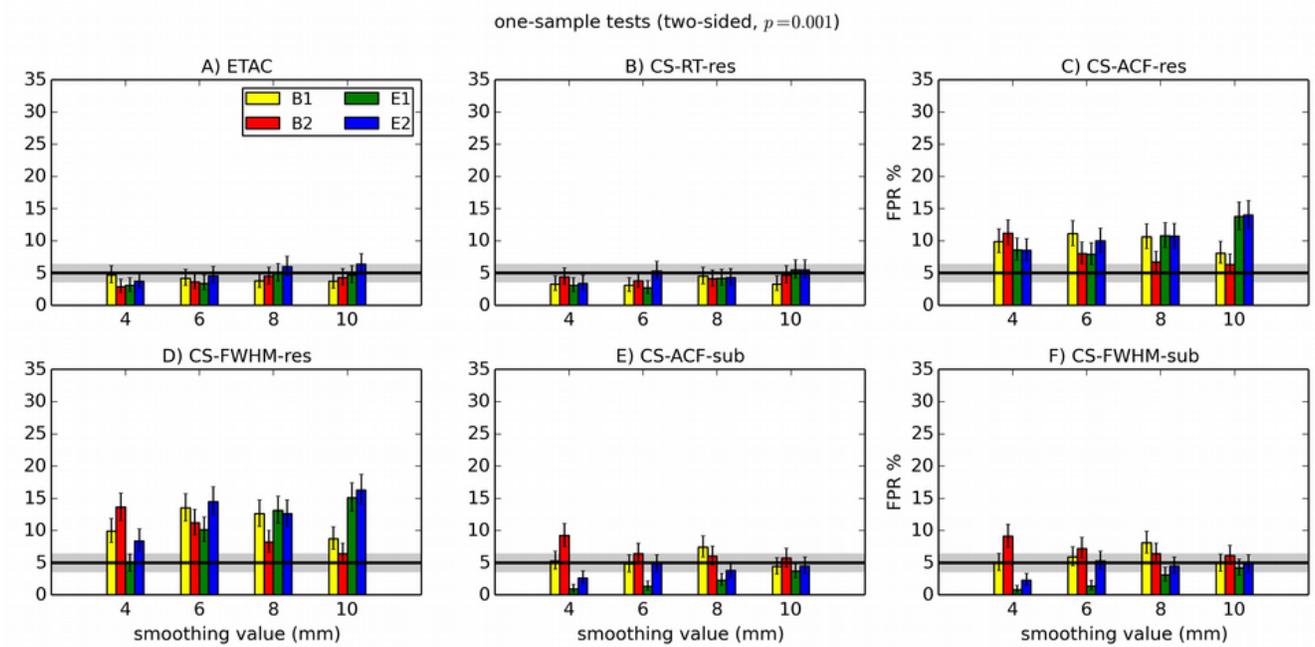

**Figure B-2**. One-sample, two-sided *t*-tests with voxelwise *p*=0.001.

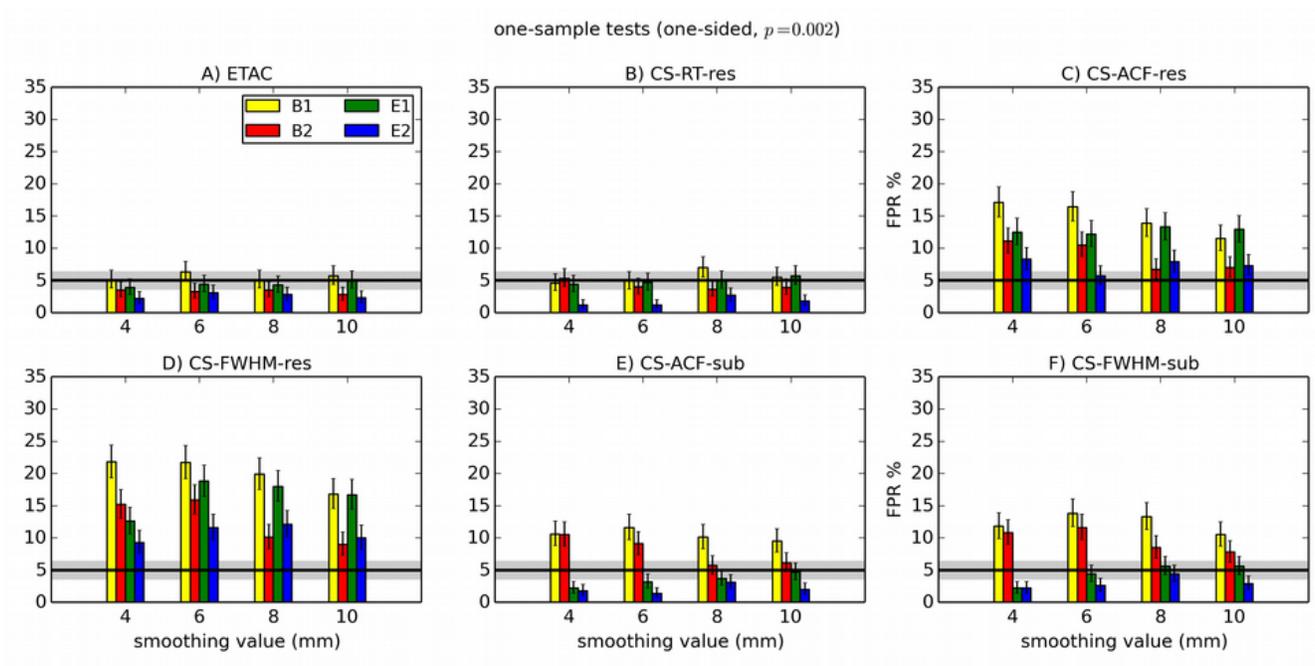

**Figure B-3**. One-sample, one-sided *t*-tests with voxelwise *p*=0.002.

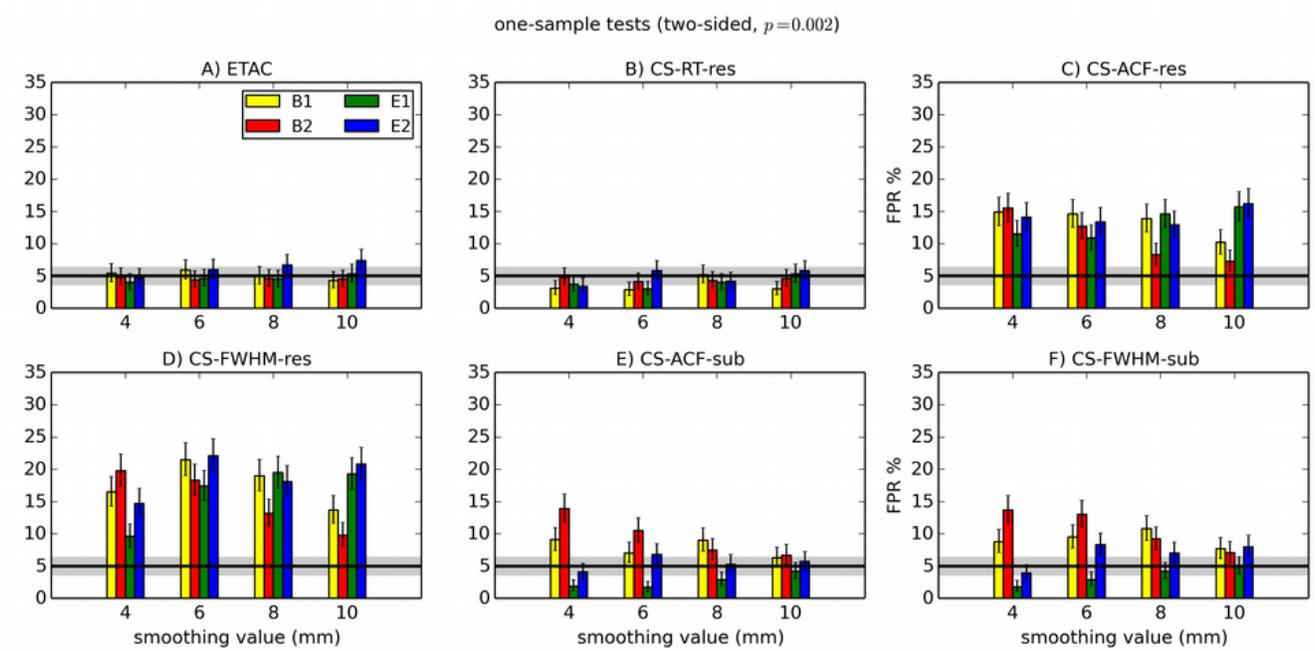

**Figure B-4**. One-sample, two-sided *t*-tests with voxelwise *p*=0.002.

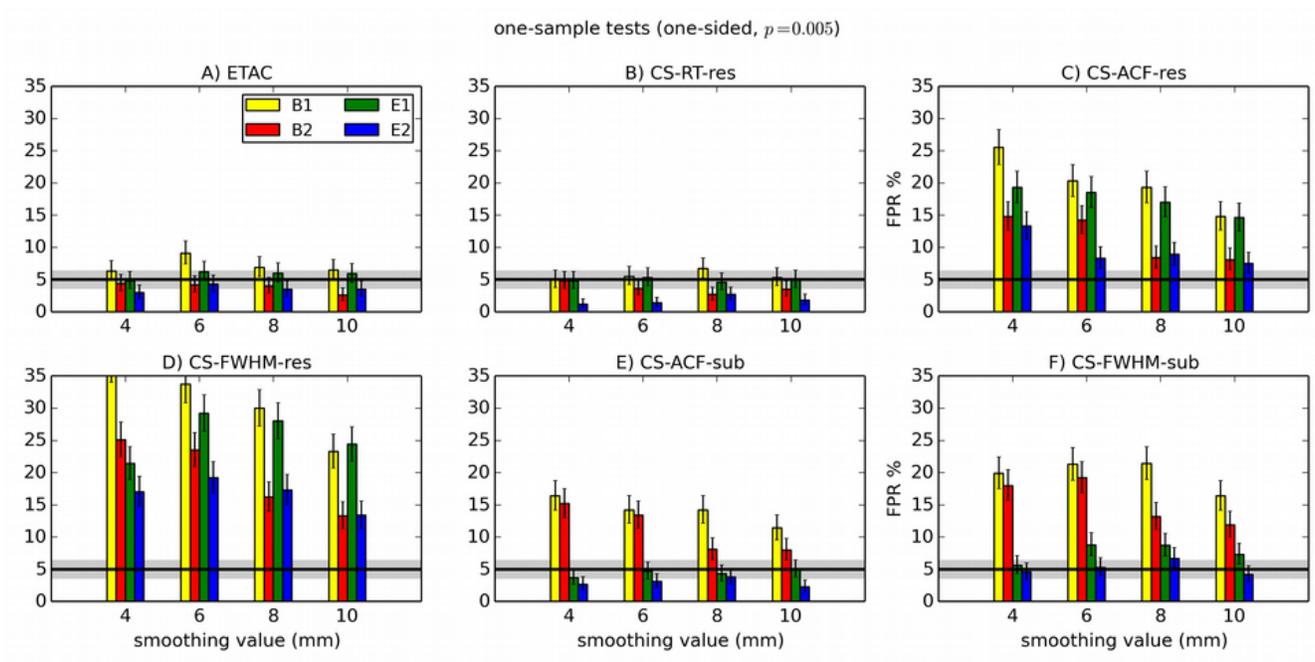

**Figure B-5**. One-sample, one-sided *t*-tests with voxelwise *p*=0.005.

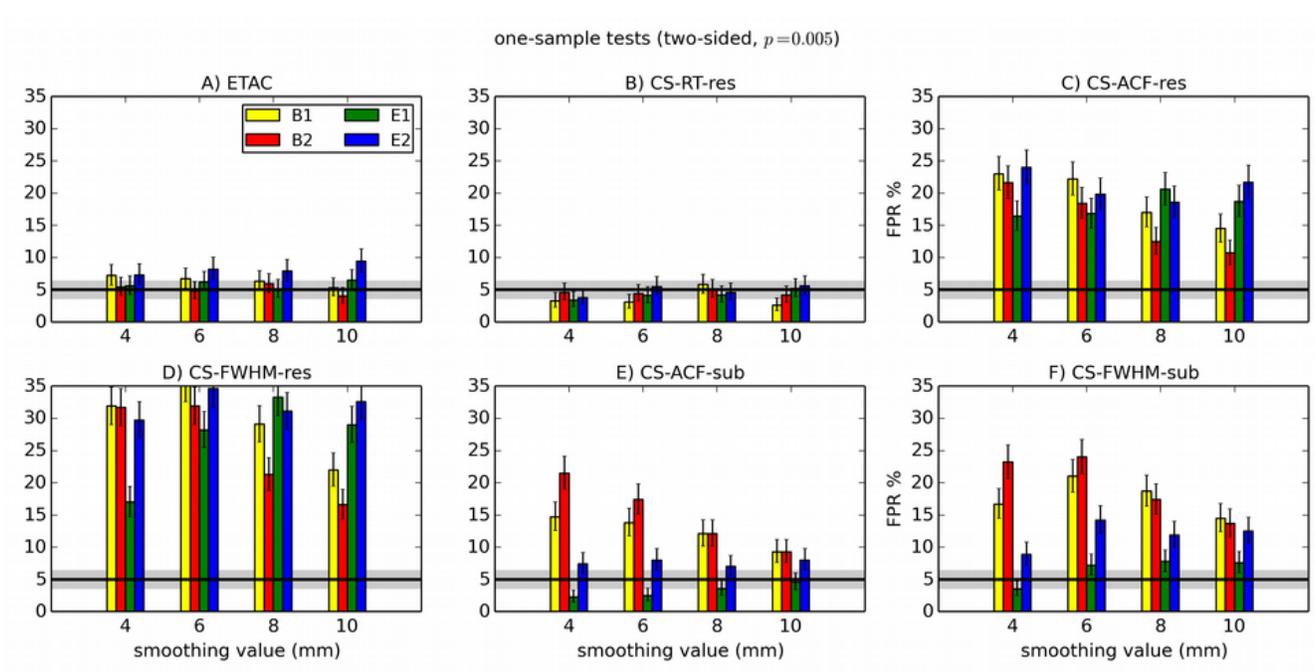

**Figure B-6**. One-sample, two-sided *t*-tests with voxelwise *p*=0.005.

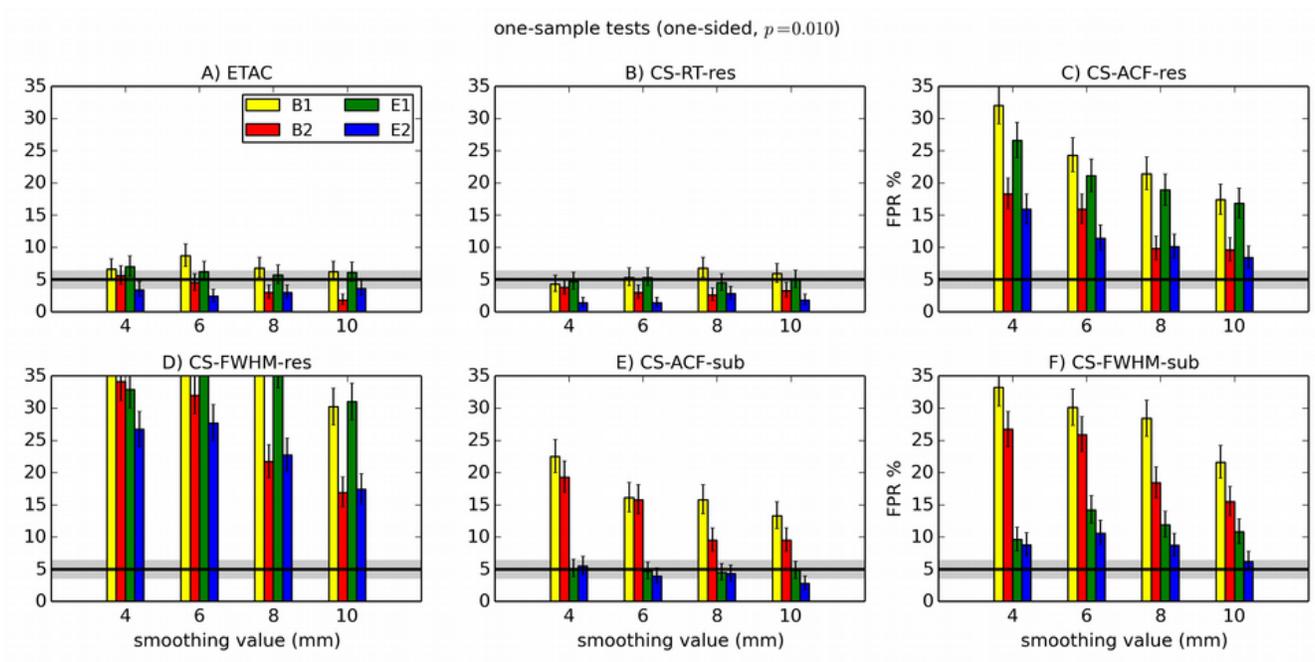

**Figure B-7**. One-sample, one-sided *t*-tests with voxelwise *p*=0.010.

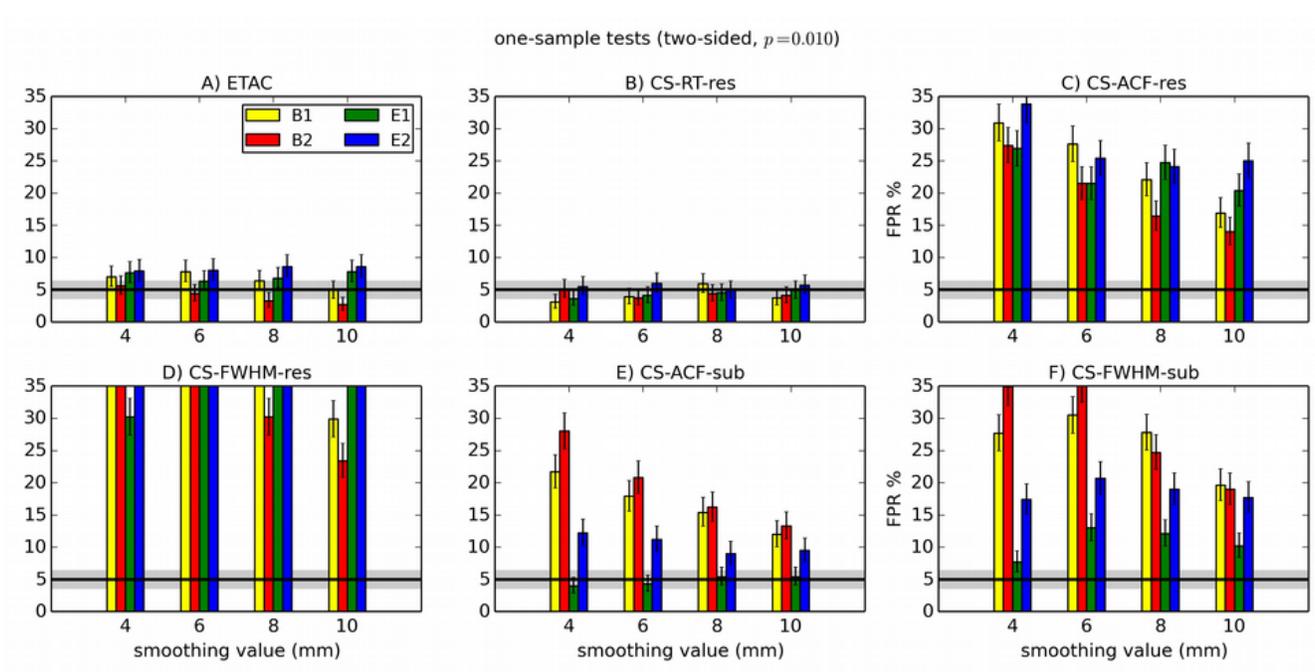

**Figure B-8**. One-sample, two-sided *t*-tests with voxelwise *p*=0.010.

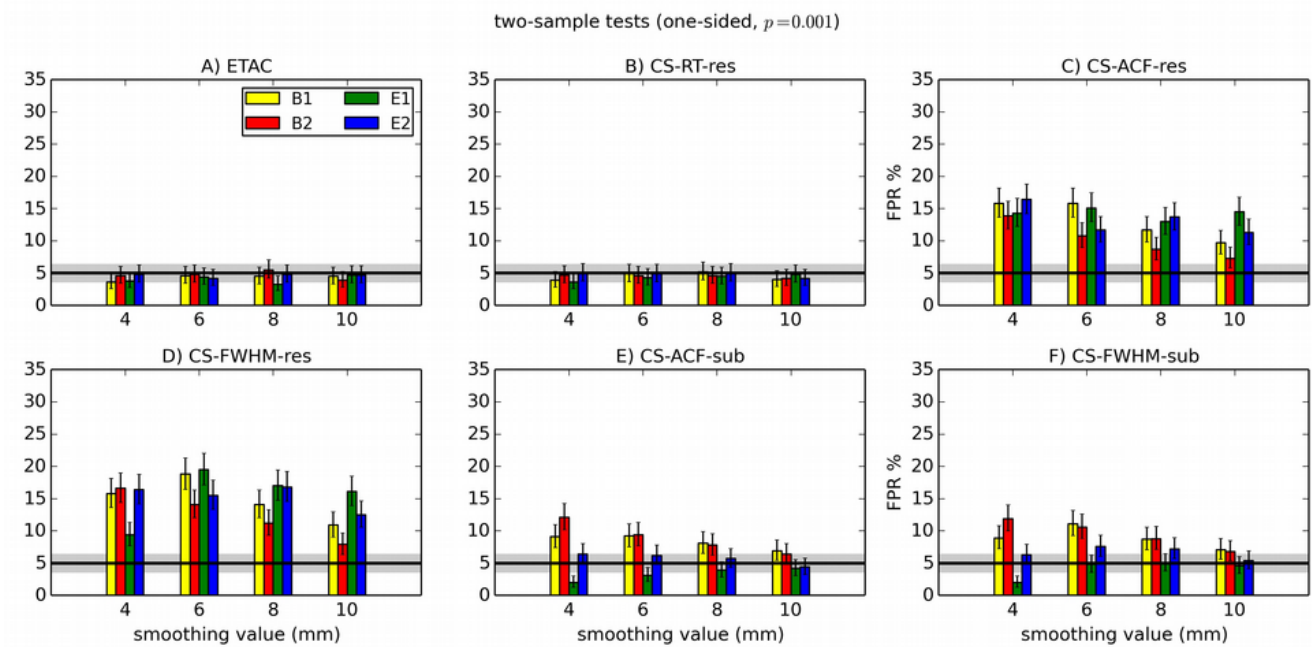

**Figure B-9**. Two-sample, one-sided *t*-tests with voxelwise *p*=0.001.

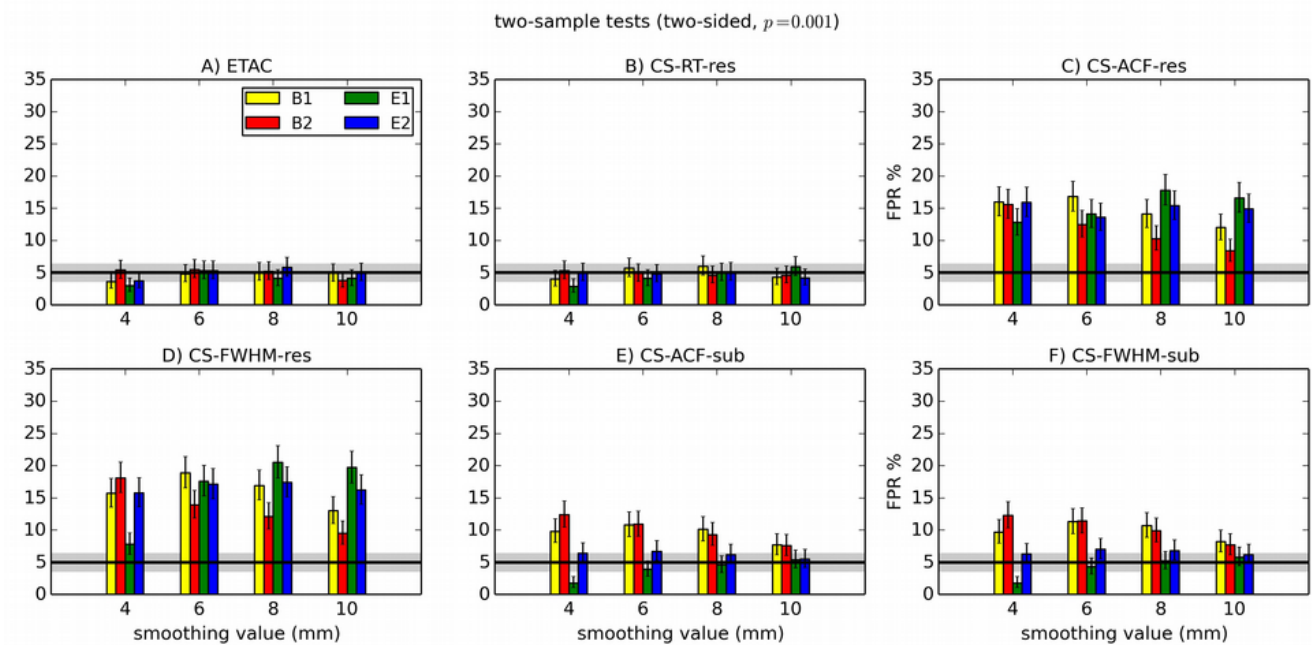

**Figure B-10.** Two-sample, two-sided *t*-tests with voxelwise *p*=0.001.

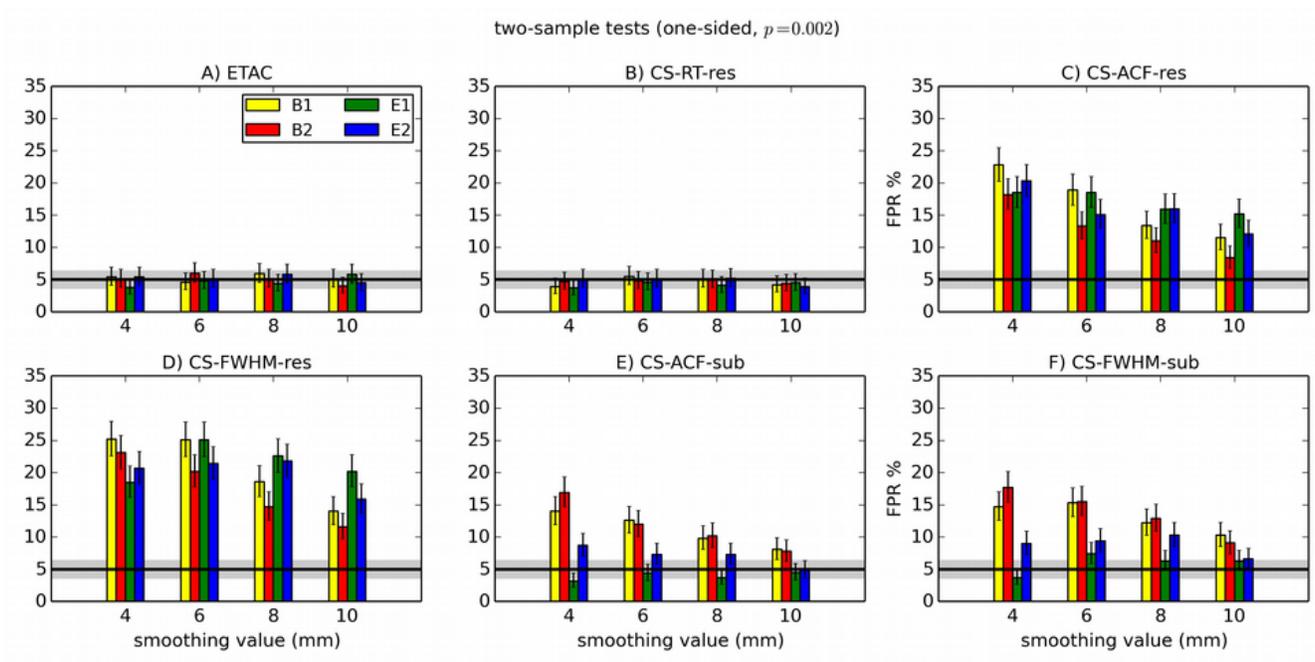

**Figure B-11**. Two-sample, one-sided *t*-tests with voxelwise *p*=0.002.

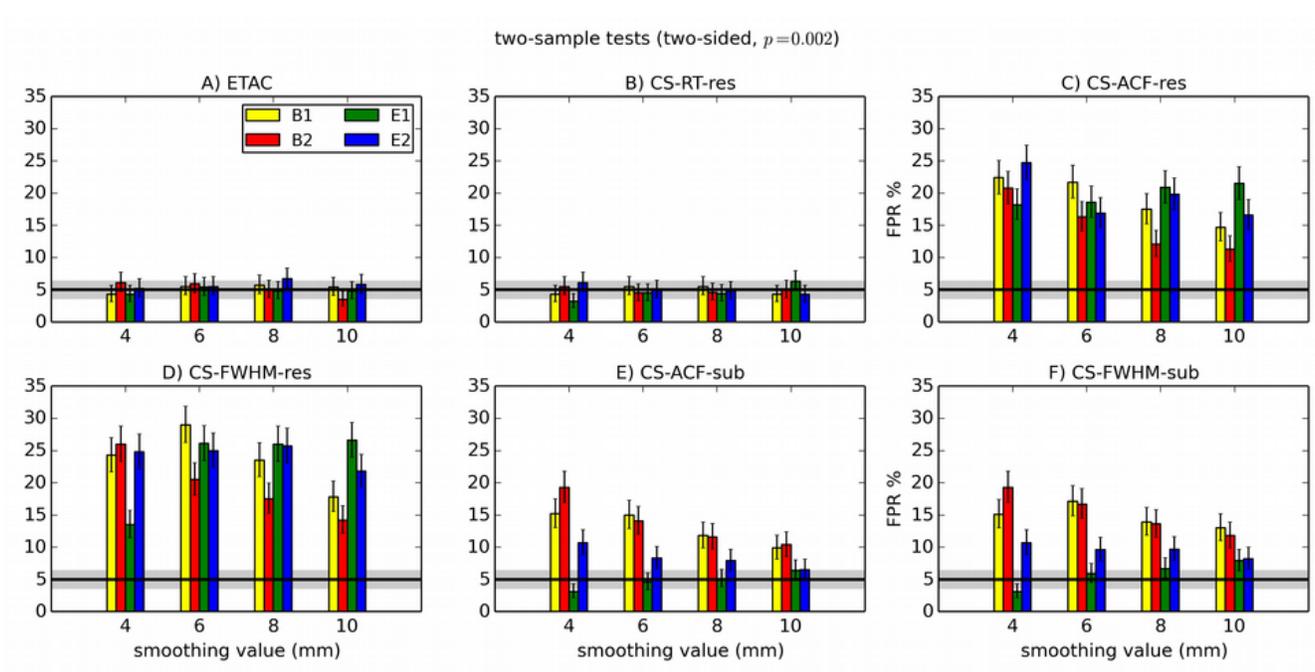

**Figure B-12**. Two-sample, two-sided *t*-tests with voxelwise *p*=0.002.

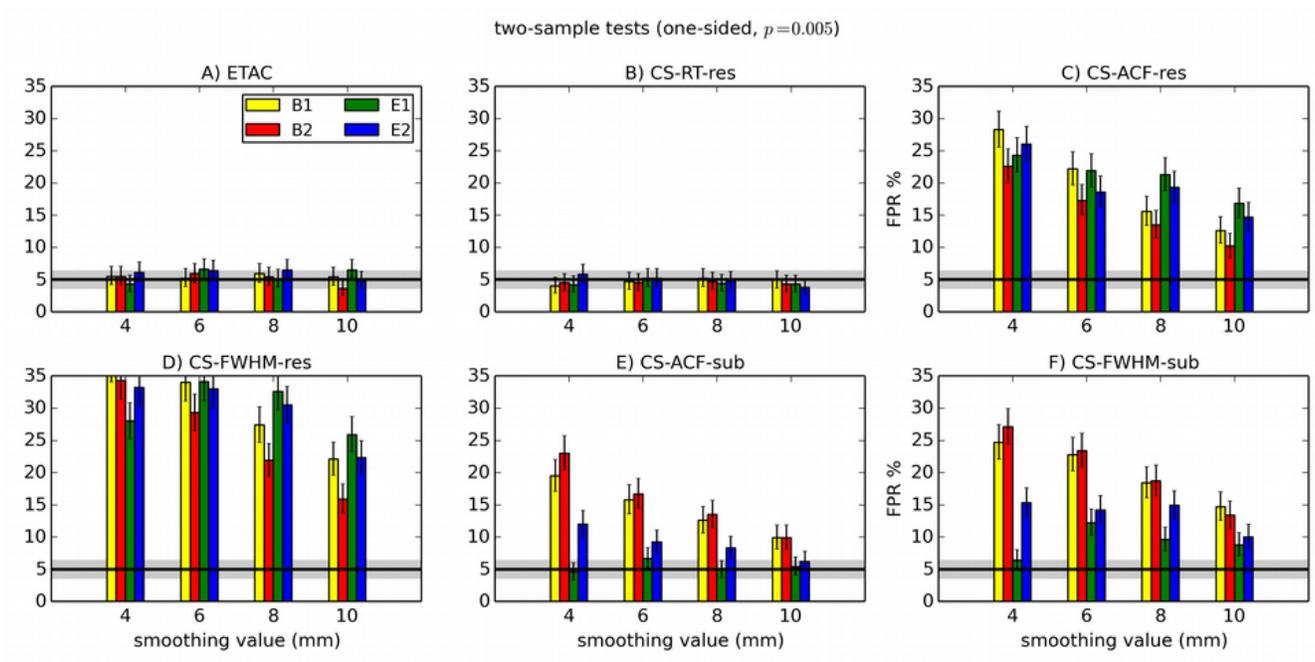

**Figure B-13**. Two-sample, one-sided *t*-tests with voxelwise *p*=0.005.

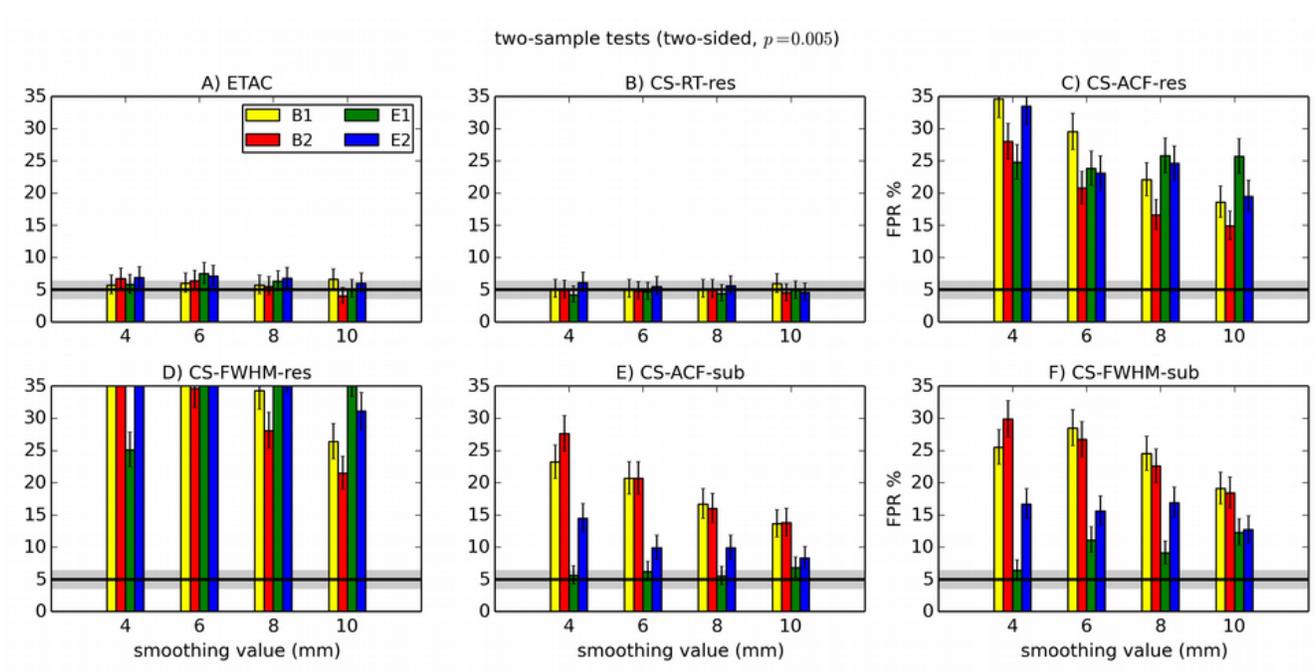

**Figure B-14**. Two-sample, two-sided *t*-tests with voxelwise *p*=0.005.

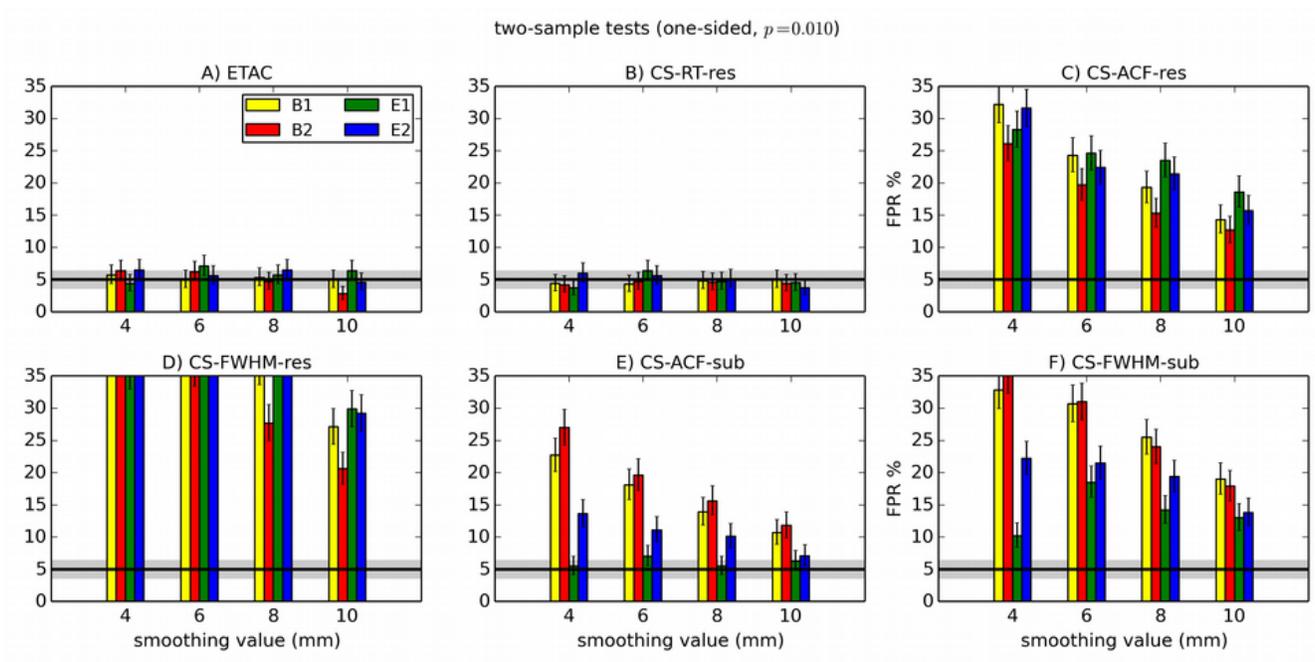

**Figure B**-15. Two-sample, one-sided *t*-tests with voxelwise *p*=0.010.

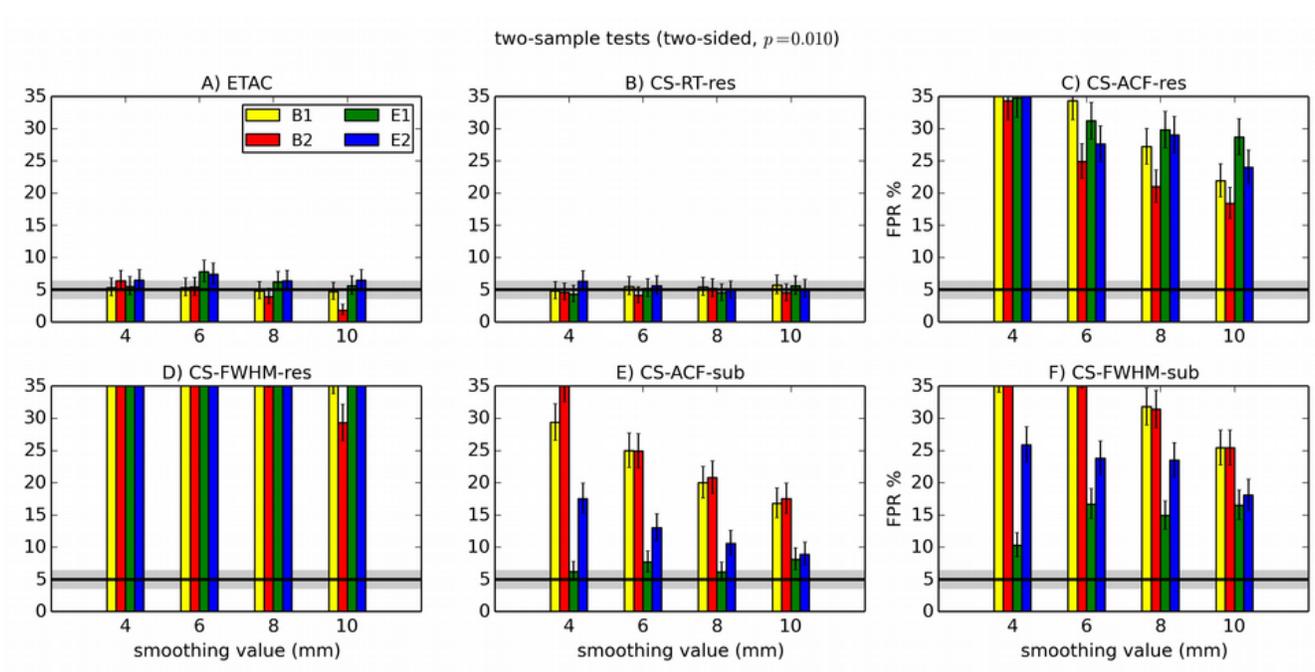

**Figure B**-16. Two-sample, two-sided *t*-tests with voxelwise *p*=0.010.

**Supplementary Information for "FMRI Clustering in AFNI: False Positive Rates Redux"**[†]

*Processing scripts*

Here we include the commands used to preprocess the data when re-running simulations for the present comparisons. First, the anatomical volume was processed and aligned to the standard space template using nonlinear warping (Script_1.warper.csh). Because many simulations were run for a given subject, this step was performed just once per subject and the results were used in each subject's afni_proc.py command, saving a large amount of time in this relatively slow step. Then, the functional data were processed using the results of the anatomical-to-standard-space alignment (Script_2B.regress.csh for block designs and Script_2E.regress.csh for event-related designs). The full processing pipeline was specified with a single afni_proc.py command (based on the help file's Example 11), which creates a full processing pipeline, given the user's choices of methods and parameters. It also generates a set of quality control scripts to view alignment, motion estimates, etc. The group analyses were carried out with a separate scripts (which will be made available online along with the data used, for convenience), which generates a random collection from the analyzed datasets from Script_2*.regress.csh, and runs 3dttest++ and 3dClustSim to produce thresholded "activation" maps. These maps are counted to produce the FPR statistics plotted in the main text.

---

```tcsh
#!/bin/tcsh

###    Script_1.warper.csh
###    written by RW Cox (NIMH, NIH, USA)

### This script nonlinear warps one anatomical dataset, taken from the
### anat_orig directory, to the MNI 2009 nonlinear template (supplied
### with AFNI binaries), and pute the resulting files into the
### anat_warped directory.

### The only command line argument is the subject ID

set sub  = $argv[1]
set site = Beijing

# set some variables if we are running SLURM

if( $?SLURM_CPUS_PER_TASK )then
  setenv OMP_NUM_THREADS $SLURM_CPUS_PER_TASK
endif

if( $?SLURM_JOBID )then
  set tempdir = /lscratch/\$SLURM_JOBID
else
  set tempdir = .
endif

# don't log AFNI programs in ~/.afni.log
# don't try any version checks
# don't auto-compress output files

setenv AFNI_DONT_LOGFILE   YES
setenv AFNI_VERSION_CHECK  NO
setenv AFNI_COMPRESSOR     NONE

### go to data directory

# topdir = directory above this Scripts directory
set topdir  = `dirname $cwd`
# set topdir = /data/NIMH_SSCC/fcon1000.perm.test/sandbox/$site

cd $topdir/anat_orig

### create final output directories

mkdir -p $topdir/anat_warped
mkdir -p $topdir/anat_warped/snapshots

### create temporary directory to hold the work, and copy the anat
### there

mkdir -p temp_$sub
cp anat_$sub.nii.gz temp_$sub
cd temp_$sub
```

```
### process the anat dataset, using the AFNI script
### that does the warping and skull-stripping

@SSwarper anat_$sub.nii.gz $sub

# compress the output datasets

gzip -1v *.nii

### move the results to where they belong

# skull-stripped original, Q-warped dataset, and the warps

\mv -f  anatSS.${sub}.nii.gz   anatQQ.${sub}.nii.gz         \
        anatQQ.${sub}.aff12.1D anatQQ.${sub}_WARP.nii.gz  \
        $topdir/anat_warped

# snapshots for visual inspection

\mv -f *.jpg    $topdir/anat_warped/snapshots

# delete the temporary directory

cd ..
\rm -rf temp_$sub

time
exit 0
```

```tcsh
#!/bin/tcsh

###   Script_2B.regress.csh
###   written by RW Cox (NIMH, NIH, USA)

### This script analyzes one subject's timeseries data (block design)
### It must be run after Script_1.warper.csh

# argv[1] = stim filename      [e.g., blk10stim.Beijing.txt]
# argv[2] = stim response      [e.g., 'BLOCK(10,1)']
# argv[3] = blur radius        [e.g., 4, 6, 8, 10]
# argv[4] = subject ID to run  [e.g., sub00156]

if( $#argv < 4 )then
  echo "Need at least 4 args"
  exit 1
endif

# set some variables if we are running SLURM

if( $?SLURM_CPUS_PER_TASK )then
  setenv OMP_NUM_THREADS $SLURM_CPUS_PER_TASK
endif

if( $?SLURM_JOBID )then
  set tempdir = /lscratch/\$SLURM_JOBID
else
  set tempdir = .
endif

# don't log AFNI programs in ~/.afni.log
# don't try any version checks
# don't auto-compress output files

setenv AFNI_DONT_LOGFILE   YES
setenv AFNI_VERSION_CHECK  NO
setenv AFNI_COMPRESSOR     NONE

# set the base dataset for MNI-izing
# (not actually needed since this was done in Script_1.warper.csh)

set basedset = MNI152_2009_template.nii.gz
set tpath = `@FindAfniDsetPath $basedset`
if( "$tpath" == '' ) then
  echo "***** @SSwarper -- Failed to find template $basedset -- exiting :("
  exit 1
endif
set basedset = $tpath/$basedset

### set the data directories we will use

# topdir = directory above this Scripts directory

set topdir  = `dirname $cwd`
```

```
set restdir = $topdir/rest_orig
set anatdir = $topdir/anat_orig
set warpdir = $topdir/anat_warped
set stimdir = $topdir/stimfiles

set start_directory = $cwd

# extract arg 1 and check it for existence

set stimfile = $argv[1]
if( ! -f $stimfile ) set stimfile = $stimdir/$stimfile
if( ! -f $stimfile ) then
  echo "argv[1] = $stimfile DOES NOT EXIST"
  exit 1
endif

# set the 'name' (label) for this stimulus in the output files

set stimfname = `basename $stimfile`
set stimname  = `basename $stimfile .txt`
set stimname  = `basename $stimname .1D`
set stimname  = `basename $stimname .Beijing`
set stimname  = `basename $stimname .Cambridge`

# extract arg 2 (HRF model)

set stimresp = $argv[2]

# extract arg 3 (blurring in mm)

set blur =     $argv[3]
set blurname = b${blur}mm

# set subject ID from arg 4

set subj = $argv[4]

# create output directory, if needed

set outdir = $topdir/$stimname.$blurname
mkdir -p $outdir
echo
echo "===== Output directory is $outdir"
echo

# copy stimfile there

if( ! -f $outdir/$stimfname )then
  cp -f $stimfile $outdir/$stimfname
endif

cd $outdir
mkdir -p snapshots
```

```
# Process this one subject

set rest_dset = $restdir/rest_${subj}.nii.gz
set anat_dset = $warpdir/anatSS.${subj}.nii.gz

# if we ran it before but it failed for some reason,
# kill the old version of the results and try again

   if( -d ${subj}.results )then
     if( ! -f ${subj}.results/stats.${subj}_REML+tlrc.HEAD )then
       \rm -rf *${subj}*
     endif
   endif

# Run this if the results don't exist and the data does

   if( ! -d ${subj}.results && -f $anat_dset && -f $rest_dset )then

     echo "-------------------------------"
     echo "Processing $subj"
     echo "-------------------------------"

# run afni_proc.py to create a single subject processing script

     afni_proc.py -subj_id $subj                                  \
         -script proc.$subj    -scr_overwrite                     \
         -blocks despike tshift align tlrc volreg                 \
                 blur mask scale regress                          \
         -copy_anat $anat_dset                                    \
            -anat_has_skull no                                    \
         -dsets $rest_dset                                        \
         -tcat_remove_first_trs 0                                 \
         -align_opts_aea -giant_move                              \
             -cost lpc+ZZ                                         \
         -volreg_align_to MIN_OUTLIER                             \
         -volreg_align_e2a                                        \
         -volreg_tlrc_warp                                        \
         -tlrc_base $basedset                                     \
         -tlrc_NL_warp                                            \
         -tlrc_NL_warped_dsets                                    \
             $warpdir/anatQQ.${subj}.nii.gz                       \
             $warpdir/anatQQ.${subj}.aff12.1D                     \
             $warpdir/anatQQ.${subj}_WARP.nii.gz                  \
         -volreg_warp_dxyz 3                                      \
         -blur_size $blur                                         \
         -regress_anaticor_fast                                   \
         -regress_anaticor_fwhm 20                                \
         -regress_stim_times $stimfile                            \
         -regress_stim_labels $stimname                           \
         -regress_basis "$stimresp"                               \
         -regress_censor_motion 0.2                               \
         -regress_censor_outliers 0.02                            \
         -regress_3dD_stop                                        \
```

```
                -regress_make_ideal_sum sum_ideal.1D                \
                -regress_est_blur_errts                             \
                -regress_reml_exec                                  \
                -regress_run_clustsim no

         # Run analysis
         tcsh -xef proc.${subj} >& proc.${subj}.output

         # If it worked, run the volreg snapshots and compress outputs
         if( -d ${subj}.results )then
            cd ${subj}.results
            @snapshot_volreg anat_final.${subj}+tlrc.HEAD             \
                  pb0?.${subj}.r01.volreg+tlrc.HEAD ${subj}
            if( -f ${subj}.jpg ) \mv -f ${subj}.jpg ../snapshots/
            gzip -1v *.BRIK *.nii
            cd ..
         endif
      else
         echo "----------------------------------------------------------------"
         echo "Skipping $subj"
         if( -d ${subj}.results ) echo "  ${subj}.results EXISTS"
         if( ! -f $anat_dset    ) echo "  $anat_dset DOES NOT EXIST"
         if( ! -f $rest_dset    ) echo "  $rest_dset DOES NOT EXIST"
         echo "----------------------------------------------------------------"
      endif

echo
echo "===== Finished ====="
echo

time
exit 0
```

```tcsh
#!/bin/tcsh

###    Script_2E.regress.csh
###    written by RW Cox (NIMH, NIH, USA)

### This script analyzes one subject's timeseries data (event design)
### It must be run after Script_1.warper.csh

# argv[1] = stim filename      [e.g., evregstim.Beijing.txt]
# argv[2] = stim response      [e.g., 'dmBLOCK']
# argv[3] = blur radius        [e.g., 4, 6, 8, 10]
# argv[4] = subject ID to run  [e.g., sub00156]

if( $#argv < 4 )then
  echo "Need at least 4 args"
  exit 1
endif

# set some variables if we are running SLURM

if( $?SLURM_CPUS_PER_TASK )then
  setenv OMP_NUM_THREADS $SLURM_CPUS_PER_TASK
endif

if( $?SLURM_JOBID )then
  set tempdir = /lscratch/\$SLURM_JOBID
else
  set tempdir = .
endif

# don't log AFNI programs in ~/.afni.log
# don't try any version checks
# don't auto-compress output files

setenv AFNI_DONT_LOGFILE   YES
setenv AFNI_VERSION_CHECK  NO
setenv AFNI_COMPRESSOR     NONE

# set the base dataset for MNI-izing
# (not actually needed since this was done in Script_1.warper.csh)

set basedset = MNI152_2009_template.nii.gz
set tpath = `@FindAfniDsetPath $basedset`
if( "$tpath" == '' ) then
  echo "***** @SSwarper -- Failed to find template $basedset -- exiting :("
  exit 1
endif
set basedset = $tpath/$basedset

### set the data directories we will use

# topdir = directory above this Scripts directory

set topdir  = `dirname $cwd`
```

```
set restdir = $topdir/rest_orig
set anatdir = $topdir/anat_orig
set warpdir = $topdir/anat_warped
set stimdir = $topdir/stimfiles

set start_directory = $cwd

# extract arg 1 and check it for existence

set stimfile = $argv[1]
if( ! -f $stimfile ) set stimfile = $stimdir/$stimfile
if( ! -f $stimfile ) then
  echo "argv[1] = $stimfile DOES NOT EXIST"
  exit 1
endif

# set the 'name' (label) for this stimulus in the output files

set stimfname = `basename $stimfile`
set stimname  = `basename $stimfile .txt`
set stimname  = `basename $stimname .1D`
set stimname  = `basename $stimname .Beijing`
set stimname  = `basename $stimname .Cambridge`

# extract arg 2 (HRF model)

set stimresp = $argv[2]

# extract arg 3 (blurring in mm)

set blur =     $argv[3]
set blurname = b${blur}mm

# set subject ID from arg 4

set subj = $argv[4]

# create output directory, if needed

set outdir = $topdir/$stimname.$blurname
mkdir -p $outdir
echo
echo "===== Output directory is $outdir"
echo

# copy stimfile there

if( ! -f $outdir/$stimfname )then
  cp -f $stimfile $outdir/$stimfname
endif

cd $outdir
mkdir -p snapshots
```

```
# Process this one subject

set rest_dset = $restdir/rest_${subj}.nii.gz
set anat_dset = $warpdir/anatSS.${subj}.nii.gz

# if we ran it before but it failed for some reason,
# kill the old version of the results and try again

   if( -d ${subj}.results )then
     if( ! -f ${subj}.results/stats.${subj}_REML+tlrc.HEAD )then
       \rm -rf *${subj}*
     endif
   endif

# Run this if the results don't exist and the data does

   if( ! -d ${subj}.results && -f $anat_dset && -f $rest_dset )then

     echo "-------------------------------"
     echo "Processing $subj"
     echo "-------------------------------"

# run afni_proc.py to create a single subject processing script

     afni_proc.py -subj_id $subj                                  \
         -script proc.$subj     -scr_overwrite                    \
         -blocks despike tshift align tlrc volreg                 \
                 blur mask scale regress                          \
         -copy_anat $anat_dset                                    \
            -anat_has_skull no                                    \
         -dsets $rest_dset                                        \
         -tcat_remove_first_trs 0                                 \
         -align_opts_aea -giant_move                              \
             -cost lpc+ZZ                                         \
         -volreg_align_to MIN_OUTLIER                             \
         -volreg_align_e2a                                        \
         -volreg_tlrc_warp                                        \
         -tlrc_base $basedset                                     \
         -tlrc_NL_warp                                            \
         -tlrc_NL_warped_dsets                                    \
               $warpdir/anatQQ.${subj}.nii.gz                     \
               $warpdir/anatQQ.${subj}.aff12.1D                   \
               $warpdir/anatQQ.${subj}_WARP.nii.gz                \
         -volreg_warp_dxyz 3                                      \
         -blur_size $blur                                         \
         -regress_anaticor_fast                                   \
         -regress_anaticor_fwhm 20                                \
         -regress_stim_times $stimfile                            \
         -regress_stim_labels $stimname                           \
         -regress_stim_types AM1                                  \
         -regress_basis "$stimresp"                               \
         -regress_censor_motion 0.2                               \
         -regress_censor_outliers 0.02                            \
```

```
        -regress_3dD_stop                                       \
        -regress_make_ideal_sum sum_ideal.1D                    \
        -regress_est_blur_errts                                 \
        -regress_reml_exec                                      \
        -regress_run_clustsim no

    # Run analysis
    tcsh -xef proc.${subj} >& proc.${subj}.output

    # If it worked, run the volreg snapshots and compress outputs
    if( -d ${subj}.results )then
      cd ${subj}.results
      @snapshot_volreg anat_final.${subj}+tlrc.HEAD           \
          pb0?.${subj}.r01.volreg+tlrc.HEAD ${subj}
      if( -f ${subj}.jpg ) \mv -f ${subj}.jpg ../snapshots/
      gzip -1v *.BRIK *.nii
      cd ..
    endif
  else
    echo "------------------------------------------------------------------"
    echo "Skipping $subj"
    if( -d ${subj}.results ) echo "  ${subj}.results EXISTS"
    if( ! -f $anat_dset    ) echo "  $anat_dset DOES NOT EXIST"
    if( ! -f $rest_dset    ) echo "  $rest_dset DOES NOT EXIST"
    echo "------------------------------------------------------------------"
  endif

echo
echo "===== Finished ====="
echo

time
exit 0
```

*Data used in Figures 1 and 4 of the main text.*

Supplementary Table 1 provides the FPR values (as decimals) for the simulation results in Fig. 1 of the main text. Supplementary Table 2 provides the FPR values (as decimals) for the simulation results shown in Fig. 5 of the main text.

**Supplementary Table 1.** FPRs for various software scenarios, with 1000 two-sample *t*-tests (as in [1,2]) using 20 subjects' data in each sample. "buggy" and "fixed" means the cluster-size thresholds selected using the Gaussian shape model with the FWHM being the median of the 40 individual subject's values; "buggy" is using 3dClustSim before the bug fix, "fixed" is using 3dClustSim after the bug fix, "mixed ACF" means the cluster-size threshold selected using Eq. (1) model for spatial correlation of the noise, with the *a,b,c* parameters being the median of the 40 individual subject's values (estimated via program 3dFWHMx). Two different per-voxel *p*-value thresholds (one-sided tests, as used in [2]) are shown. The 95% confidence interval for the expected FPR=0.05 out of 1000 trials is 0.036-0.064.

| blur | stim | buggy $p=0.01$ | fixed $p=0.01$ | mixed ACF $p=0.01$ | buggy $p=0.001$ | fixed $p=0.001$ | mixed ACF $p=0.001$ |
|---|---|---|---|---|---|---|---|
| 4 mm | B1 | 0.384 | 0.355 | 0.274 | 0.117 | 0.111 | 0.109 |
| 6 mm | B1 | 0.406 | 0.358 | 0.203 | 0.149 | 0.123 | 0.097 |
| 8 mm | B1 | 0.367 | 0.346 | 0.276 | 0.125 | 0.114 | 0.113 |
| 10 mm | B1 | 0.321 | 0.272 | 0.137 | 0.125 | 0.108 | 0.087 |
| 4 mm | B2 | 0.367 | 0.346 | 0.276 | 0.125 | 0.114 | 0.113 |
| 6 mm | B2 | 0.352 | 0.322 | 0.192 | 0.124 | 0.109 | 0.098 |
| 8 mm | B2 | 0.317 | 0.260 | 0.162 | 0.112 | 0.096 | 0.081 |
| 10 mm | B2 | 0.250 | 0.222 | 0.136 | 0.097 | 0.083 | 0.064 |
| 4 mm | E1 | 0.150 | 0.136 | 0.071 | 0.033 | 0.030 | 0.028 |
| 6 mm | E1 | 0.217 | 0.173 | 0.071 | 0.079 | 0.064 | 0.042 |
| 8 mm | E1 | 0.242 | 0.178 | 0.078 | 0.100 | 0.071 | 0.048 |
| 10 mm | E1 | 0.231 | 0.181 | 0.078 | 0.106 | 0.075 | 0.049 |
| 4 mm | E2 | 0.238 | 0.212 | 0.125 | 0.069 | 0.062 | 0.057 |
| 6 mm | E2 | 0.257 | 0.240 | 0.101 | 0.099 | 0.074 | 0.059 |
| 8 mm | E2 | 0.288 | 0.232 | 0.097 | 0.103 | 0.075 | 0.059 |
| 10 mm | E2 | 0.259 | 0.215 | 0.092 | 0.100 | 0.078 | 0.059 |

**Supplementary Table 2.** FPRs for various software scenarios, analogous to Table 1 but with cluster-size thresholds now determined from the '-Clustsim' option of 3dttest++ (one-sided tests with NN=1 nearest neighbor clustering). Results from the five other possible combination of sidedness and neighborhoods are very similar. NB: fractional FPR values are shown here, not percentages as in the text plots.

| blur  | stim | *p*=0.01 | *p*=0.007 | *p*=0.005 | *p*=0.003 | *p*=0.002 | *p*=0.0015 | *p*=0.001 |
|-------|------|----------|-----------|-----------|-----------|-----------|------------|-----------|
| 4 mm  | B1   | 0.048    | 0.050     | 0.048     | 0.050     | 0.047     | 0.046      | 0.042     |
| 6 mm  | B1   | 0.045    | 0.043     | 0.047     | 0.050     | 0.049     | 0.045      | 0.046     |
| 8 mm  | B1   | 0.051    | 0.051     | 0.050     | 0.046     | 0.046     | 0.044      | 0.044     |
| 10 mm | B1   | 0.044    | 0.051     | 0.056     | 0.056     | 0.052     | 0.052      | 0.046     |
| 4 mm  | B2   | 0.048    | 0.055     | 0.045     | 0.049     | 0.052     | 0.053      | 0.046     |
| 6 mm  | B2   | 0.048    | 0.046     | 0.050     | 0.052     | 0.049     | 0.050      | 0.047     |
| 8 mm  | B2   | 0.051    | 0.050     | 0.050     | 0.050     | 0.048     | 0.053      | 0.052     |
| 10 mm | B2   | 0.047    | 0.045     | 0.051     | 0.047     | 0.045     | 0.043      | 0.047     |
| 4 mm  | E1   | 0.048    | 0.049     | 0.049     | 0.050     | 0.045     | 0.040      | 0.038     |
| 6 mm  | E1   | 0.051    | 0.053     | 0.051     | 0.046     | 0.051     | 0.054      | 0.049     |
| 8 mm  | E1   | 0.050    | 0.049     | 0.048     | 0.051     | 0.052     | 0.049      | 0.050     |
| 10 mm | E1   | 0.057    | 0.057     | 0.055     | 0.051     | 0.053     | 0.050      | 0.049     |
| 4 mm  | E2   | 0.042    | 0.044     | 0.048     | 0.043     | 0.042     | 0.038      | 0.036     |
| 6 mm  | E2   | 0.044    | 0.048     | 0.042     | 0.044     | 0.041     | 0.043      | 0.039     |
| 8 mm  | E2   | 0.048    | 0.048     | 0.045     | 0.050     | 0.044     | 0.047      | 0.045     |
| 10 mm | E2   | 0.048    | 0.048     | 0.049     | 0.048     | 0.046     | 0.048      | 0.046     |